# Investigating the Impacts of AGN Activities on Dwarf Galaxies with FAST H I Observations

Hong-Ying Chen[1], Chao-Wei Tsai[1,2,3]⋆, Pei Zuo[1], Niankun Yu[4,1], Jialai Wang[5,6,1],
Kai Zhang[1], Guodong Li[7], Yogesh Chandola[8,9], Zheng Zheng[1], Jingwen Wu[3,1], Di Li[10,1], Lulu Bao[3]

[1] *National Astronomical Observatories, Chinese Academy of Sciences, A20 Datun Road, Chaoyang District, Beijing, 100101, China*
[2] *Institute for Frontiers in Astronomy and Astrophysics, Beijing Normal University, Beijing 102206, China*
[3] *University of Chinese Academy of Sciences, 19A Yuquanlu, Beijing, 100049, China*
[4] *Max Planck Institute for Radio Astronomy, Auf dem Hügel 69, 53121 Bonn, Germany*
[5] *Department of Astronomy, University of Science and Technology of China, Hefei 230026, China*
[6] *School of Astronomy and Space Science, University of Science and Technology of China, Hefei 230026, China*
[7] *Kavli Institute for Astronomy and Astrophysics, Peking University, Beijing 100871, People's Republic of China*
[8] *Indian Institute of Astrophysics (IIA), 2nd Block, Koramangala, Bengaluru, 560034, India*
[9] *Purple Mountain Observatory, Chinese Academy of Sciences, 10 Yuan Hua Road, Qixia District, Nanjing 210023, Peoples Republic of China*
[10] *New Cornerstone Science Laboratory, Department of Astronomy, Tsinghua University, Beijing 100084, China*

**ABSTRACT**
We present the results of H I line observations towards 26 Active Galactive Nuclei (AGN)-hosting and one star-forming dwarf galaxies ($M_* < 10^{9.5}\ M_\odot$) with the 19-beam spectral line receiver of FAST at 1.4 GHz. Our FAST observed targets are combined with other AGN-hosting dwarf galaxies covered in the ALFALFA footprint to form a more comprehensive sample. Utilizing the information from optical surveys, we further divide them into isolated and accompanied subsamples by their vicinity of nearby massive galaxies. We compare the H I gas abundance and star-forming rate (SFR) between the subsamples to assess the role of internal and external processes that may regulate the gas content in dwarf galaxies. As a result, we find that AGN are more commonly identified in accompanied dwarf galaxies than in their isolated counterparts. Meanwhile, AGN-hosting dwarf galaxies have slightly but significant lower H I mass fraction relatively to the non-AGN control sample in accompanied dwarf galaxies. On the other hand, we find a decreasing SFR in AGN-hosting dwarf galaxies towards denser environments, as well as an extremely low incidence of quenched isolated dwarfs within both AGN and non-AGN subsamples. These results indicate that although these AGN could potentially regulate the gas reservoir of dwarf galaxies, environmental effects are likely the dominant quenching mechanism in the low-mass universe.

## 1 INTRODUCTION

In massive galaxies, the correlation between the mass of super massive black holes (SMBHs) and physical properties of their hosts is well-established in current Λ Cold Dark Matter (ΛCDM) models (e.g. Kormendy & Ho 2013; Ding et al. 2020). Dwarf galaxies ($M_* < 10^{9.5}\ M_\odot$); Reines et al. 2013 with shallow gravitational well, lower metallicity and mass-to-light ratio, provide an ideal test bed for the ΛCDM models. In fact, many observational evidences fail to fit into the predictions of the standard model in the low-mass universe, such as lesser observed satellite dwarfs as predicted (the "missing satellites" problem, Moore et al. 1999), absence of cuspy dark matter profiles (Oh et al. 2011), and the rarity of most massive dwarfs (the "too big to fail" problem, Boylan-Kolchin et al. 2011). The causes of these unresolved issues are still under debate, yet most likely either a modified cosmological dark matter model, or reasonable mechanisms of baryonic mass depletion is required. The latter are usually considered in the context of supernova feedback (Anglés-Alcázar et al. 2017; Habouzit et al. 2017), reionization suppression (Brown et al. 2014; Weisz et al. 2014; Skillman et al. 2017; Bettinelli et al. 2018), and ram pressure/tidal stripping (Gunn & Gott 1972a). Although AGN activity in the dwarf galaxies is usually mild or weak (Mezcua 2017; Reines et al. 2013; Moran et al. 2014), its contribution could be still significant in removing the gas from the low-mass galaxy systems. To further investigate this discrepancy between observational results and theoretical models for a more comprehensive understanding on the galaxy-BH co-evolution in the low-mass regime, we aim to evaluate the role of the feedback mechanisms by AGN.

Star formation suppression in galaxies may come from external environment or internal processes such as stellar and active galactic nuclei (AGNs) feedback. Peng et al. (2010, 2012) demonstrate that internal and external quenching avenues are largely independent to each other. Massive galaxies are capable of maintaining a hot gaseous halo which eventually cease the star formation by intrinsic feedback from stellar and AGN activity (e.g. Birnboim & Dekel 2003; Kereš et al. 2005; Croton 2006; Dekel & Birnboim 2006, 2008). However, in low-mass galaxies ($M_* < 10^{10}\,M_\odot$) with weak or no associated hot halos, cold gas that temporarily ejected by stellar feedback could eventually cool down and be recycled (e.g. Dekel & Silk 1986; Stinson et al. 2007; Bradford et al. 2015). On the other hand, because of the generally smaller black hole mass (Volonteri et al. 2008) to drive strong feedback, galaxies in the mass range of dwarfs were thought to quench primarily via environmental effects.

⋆ cwtsai@nao.cas.cn

External processes such as ram pressure stripping (Gunn & Gott 1972a), harassment, and strangulation (Larson et al. 1980) can strip the ambient gas of dwarfs, preventing further gas accreting, and ultimately shut off their star formation. In fact, quenched dwarf galaxies are thought to be exceedingly rare in the field (Geha et al. 2012). Additionally, certain observations indicate that some galaxies exhibit features of suppressions due to reionization reflected by their star formation histories (Brown et al. 2014; Weisz et al. 2014; Skillman et al. 2017; Bettinelli et al. 2018), while various others show no definite signature (Grebel & Gallagher 2004; Monelli et al. 2010; Hidalgo et al. 2011).

AGN feedback powered by intermediate mass black hole (IMBH; $M_{BH} \sim 10^4 - 10^6\,M_\odot$) accretion in dwarf galaxies was revealed to be more effective than the prediction in current cosmological simulations (e.g. Reines & Comastri 2016; Moran et al. 2014; Sartori et al. 2015; Mezcua 2017; Greene et al. 2020). AGN feedback is often considered in the theoretical models of low-mass galaxy evolution (e.g. Manzano-King et al. 2019; Dashyan et al. 2018; Koudmani et al. 2019; Zhang et al. 2020; Cai et al. 2020, 2021). But yet, results from some other studies show evidence opposing the importance of AGN feedback in dwarf systems. For example, Trebitsch et al. (2018) argued that AGN feedback is negligible compared to stellar activity in such systems based on high-resolution cosmological simulations. Because of these contradicting results, it is crucial to better understand what fraction of dwarf galaxies host an AGN, how powerful these AGN can be and whether AGN feedback is the dominant contribution process to regulate the star formation in these dwarf galaxies. To answer these questions, statistical analysis based on a large sample is required in order to quantitatively assess the role of AGN in the gas environment of dwarf galaxies.

The 21 cm hyperfine emission line produced by neutral atomic hydrogen (H I) is the most important tracer of the interstellar medium (ISM) associated with galaxies with various morphological types at different evolution stages. The observations of H I emission provide an ideal probe of both the structural and dynamical information of the AGN activity in dwarf galaxies, such as gas velocity dispersion, neutral gas abundance, baryonic mass, etc. These pieces of information are crucial for studying quenching mechanisms within dwarf galaxies. H I line was used to diagnose the impact of AGN in galaxies (e.g. Chandola et al. 2020; Maccagni et al. 2017; Morganti et al. 2005, 1998; Ho et al. 2008). Ho et al. (2008) suggests that the massive host galaxies of type I AGN appeared to be more gas-rich than the inactive objects of the same morphological type. Other recent studies show that the H I gas mass to stellar masses ratio is more strongly correlated with black hole masses than with stellar mass (Wang et al. 2024). These works demonstrate how H I measurements can contribute to our understanding of dwarf galaxy evolution.

In this paper, we present the results from our exploratory H I observations on a sample of 27 low-$z$ dwarf galaxies. The primary goal of this work is to measure the hydrogen gas mass of our targets, in order to have a statistical view on the effect of AGN feedback and environmental effects on gas abundance in dwarfs.

We explain the sample selection methodology in Section 2, followed by a description of FAST (Li et al. 2018) observations and data reduction in Section 3. In Section 4 and 5, we present the results and discussions. A summary of our main arguments is given in Section 6. For this paper, we assume a flat $\Lambda$CDM universe with $H_0 = 70$ km s$^{-1}$ Mpc, $\Omega_m = 0.3$ and $\Omega_\Lambda = 0.7$.

## 2 SAMPLE SELECTION

The dwarf galaxies sample of this study comprises four subsamples. We summarize the sizes of these subsamples in Table 1 and provide a detailed description of each in this section.

### 2.1 FAST observed sample

Facilitating the Sloan Digital Sky Survey Data Releases 7 and 8 (SDSS DR7 and DR8), hundreds of local AGN-host dwarf galaxies ($z < 0.1$) are identified by means of BPT and emission-line measurements (Reines et al. 2013; Moran et al. 2014; Sartori et al. 2015). Based on these studies, Manzano-King et al. (2019) performed spatially resolved long-slit spectroscopy for 50 selected sources with the Low Resolution Imaging Spectrometer (LRIS; Oke et al. 1995; Rockosi et al. 2010) on the Keck I telescope. The 50 selected sources consist of 29 dwarfs with optical signatures of AGN activity shown in their BPT diagram, and a control sample of 21 non-AGN star-forming (SF) dwarf galaxies. AGN-driven ionized gas outflow were seen in 9 AGN and 4 SF sources indicated by spatially extended outflows. We assess the detectability of the sources by their expected H I line fluxes assuming an H I mass fraction ($f_{H\,I}= M_{H\,I}/M_*$) of $\sim 1$, which is the typical value for galaxies with $M_* \sim 10^{9.5}\,M_\odot$ (Bradford et al. 2018; Parkash et al. 2018). Sources with higher expected flux density measurements are given with higher priorities for FAST observations. To avoid contaminations from radio frequency interference (RFI), we excluded those with a redshifted central frequency with known severe RFI at FAST site. As a result, we selected 4 sources (2 AGN, 1 composite system with both AGN and SF components, and 1 non-AGN SF dwarfs) to be observed with FAST.

While the BPT diagnostic provides an efficient approach to establish a large and widely spread sample of AGN in dwarfs, detecting AGN variability features is more sensible to weaker, IMBH-powered AGN (Baldassare et al. 2020a,b; Ward et al. 2021). In the study of Baldassare et al. (2020a), 102 dwarf galaxies in NASA-Sloan Atlas[1] (NSA; Blanton & Moustakas 2009) were found to have AGN-like stochastic variability in their light curves from the monitoring by the Palomar Transient Factory (Law et al. 2009). Priority was given to those with relatively higher mass and lower redshift. Amongst the 102 galaxies, 44 have a $z_{spec} < 0.03$ and $M_* < 10^{9.5}\,M_\odot$. As described above, higher mass and lower redshift targets are preferred for FAST H I observations. We further exclude sources that are inaccessible to FAST due to its latitude constraints or lie at redshifts where the data may be heavily contaminated by RFI. As a result, 23 sources are also selected to be observed with FAST.

In total, we observed 27 dwarf galaxies (25 AGN, 1 composite, and 1 non-AGN SF) with FAST in L band. Because of FAST's effective aperture of 300 m, the angular resolution of our FAST observations is 2.9′, approximately the same as the ALFALFA's resolution. One of the variable AGN-host dwarfs (NSA153595) suffered from unknown severe RFI during the observation, we excluded this source from the final sample. Therefore, we present observed data of the rest 26 sources in this paper. In addition, we look for stellar mass information for all observed targets (see Section 2.3). One variable AGN J094419.41+095905.3 has a stellar mass of $10^{9.58}$ $M_\odot$ (see Table 2), which is beyond the dwarf galaxy regime, we do not include this source in our statistical analysis in the following sections.

---

[1] http://www.nsatlas.org

### 2.2 The $\alpha$.100 catalogue

The Arecibo Legacy Fast ALFA (ALFALFA; Haynes et al. 2018) Survey covers nearly 7000 deg$^2$ of high Galactic latitude sky. The ALFALFA survey ($\alpha$.100) consists of $\sim$ 31,500 sources at $z < 0.06$. The $\alpha$.100 sample reaches a 5$\sigma$ detection limit of $\sim$ 0.72 Jy km s$^{-1}$ for $W_{50}$ = 200 km s$^{-1}$. Within the $\sim$ 31,500 sources in the $\alpha$.100 catalog, Yu et al. (2022) compiled a list of 13,511 galaxies with supplementary stellar mass and SFR information, obtained from the second version of GALEX-SDSS-WISE Legacy Catalog (GSWLC-X2[2]; Salim et al. 2016; Salim et al. 2018). Out of the 13,511 galaxies, there are 5,552 dwarf galaxies ($M_* < 10^{9.5}$ $M_\odot$) with measured GSWLC-X2 stellar mass in $\alpha$.100. These 5,552 dwarf galaxies are used as a complementary sample to our dwarf galaxies.

### 2.3 The NMJG catalogue

In addition to the H I observed dwarf galaxies, we further expand our dwarf galaxy sample from NSA v1_0_1, the MPA-JHU[3] database, and GSWLC-X2. For simplicity, we refer this NSA-MPA-JHU-GSWLC combined sample NMJG catalogue. The NSA v1_0_1 combines GALEX imaging data with reprocessed SDSS DR8 photometry applied with an improved background subtraction technique (Blanton et al. 2011). The parent dataset contains $\sim$ 640,000 galaxies with redshifts $z < 0.055$ within the SDSS footprint. Stellar masses calculated by elliptical Petrosian aperture photometry is supplied in NSA. The MPA-JHU database contains $\sim$ 1,500,000 spectroscopic data of extragalactic sources. Stellar masses in MPA-JHU are calculated using the Bayesian methodology and model grids described in Kauffmann et al. (2003). These sources are classified into galaxy, quasar, and stars by performing a rest-frame principal-component analysis (PCA) (Bolton et al. 2012). Spectroscopic diagnosis using the Baldwin, Phillips & Terlevich (BPT) diagnostic (Baldwin et al. 1981) further divides galaxies into "Star Forming", "Composite", "AGN", "Low S/N Star Forming", "Low S/N AGN", and "Unclassifiable" categories. The GSWLC-X2 catalogue encompasses total stellar masses and SFRs measured by spectral energy distribution fitting of the global photometry from the mid-infrared, optical, and ultra-violet bands. The typical uncertainties of the stellar masses and SFRs are 0.042 dex and 0.064 dex, respectively. GSWLC contains $\sim$ 660,000 galaxies within the GALEX footprint, regardless of a UV detection, altogether covering 90% of SDSS with $0.01 < z < 0.30$. We merge the three databases, excluding sources with a MPA-JHU source type of "star". The NMJG catalogue contains 1,131,485 optical sources. In our analysis, we primarily adopt stellar masses ($M_*$) from the GSWLC-X2 catalog. For sources without a good estimation of $M_{*,\mathrm{GSWLC}}$ with $\chi_r^2 < 30$ (flag_sed $\neq$ 2), we adopted stellar mass measured by K-correction fit for Sersic fluxes in NSA when available. For the rest of the sources, we used the median estimate of the total stellar mass probability density function (PDF) derived from model photometry (lgm_tot_p50) provided by the MPA-JHU catalogue when available.

### 2.4 Optical counterparts and nearest massive neighbours

We search for optical counterparts for the dwarf galaxies in the $\alpha$.100 sample and our FAST observed targets from the NMJG catalogue. For the 22 dwarf galaxy systems with variable AGN selected from Baldassare et al. (2020a), we identified their optical data directly by the unique NSA source ID in NSA v0_1_2. For other sources, we match them with the nearest NMJG optical sources within a spatial separation of 3″, and a maximum offset between the H I and the optical velocities of 300 km s$^{-1}$. We note that only two sources (AGC 194713 and AGC 262402) are classified as "composite" consisting both AGN and starburst features by BPT diagnostics in MPA-JHU. However, due to deficient metallicity and less massive blackhole, AGN in dwarfs would possibly submerged in normal star-forming dwarf (Baldassare et al. 2020a,b; Sartori et al. 2015). The anticipated rate for the AGN/starforming mixture systems would be higher than that identified using BPT.

To expand our sample of dwarf galaxies, we include 11,989 dwarf galaxies in the NMJG catalogue covered by the ALFALFA footprint excluding those detected in $\alpha$.100 or our FAST observations. Out of the 11,989 dwarf galaxies, 313 are identified as AGN-hosts by BPT diagnostics, we include the rest 10,137 as non-AGN dwarf galaxies, we estimated the upper limits to their $M_{\rm H\,I}$ based on the 50% completeness of $\alpha$.100, which is considered the "sensitivity limit" of the survey (Haynes et al. 2011; see Section 3.2).

In addition, we investigate the environment of our sample by identifying the nearest massive neighbours with $M_*> 10^{9.5}$ $M_\odot$ in the NMJG catalogue for each FAST observed target and the full $\alpha$.100 dwarf galaxy sample. As suggested by Geha et al. (2012), less than 20% galaxies are quenched without a massive neighbour within 1.5 Mpc. Therefore, we search for massive neighbours for our sample within a projected distance $d$ of 1.5 Mpc and a velocity difference of < 500 km s$^{-1}$. To avoid mis-identifications caused by source confusion, we exclude galaxies with $d < 10$ kpc, which corresponds to the typical virial radius for dwarf galaxies. Galaxies in our sample with a nearby massive galaxy within 10 kpc − 1.5 Mpc are labeled as "accompanied galaxies", and otherwise "isolated". In total, there are 322 accompanied dwarfs with AGN (17 FAST targets, out of which 11 are detected and 6 undetected; 2 listed in Yu et al. 2022; 303 NMJG sources that were undetected in $\alpha$.100.), 15 isolated dwarf galaxies with AGN (7 observed with FAST, out of which 5 are detected and 2 undetected; 8 NMJG sources that were undetected in $\alpha$.100), 14,553 accompanied non-AGN dwarfs (including 1 detected by FAST, 4,415 listed in Yu et al. 2022, and 10,137 NMJG sources that were undetected in $\alpha$.100), and 2934 isolated non-AGN dwarf galaxies (including 1,135 listed in Yu et al. 2022 and 1,799 NMJG sources that were undetected in $\alpha$.100). We summarize the sizes of the four subsamples in Table 1.

---

[2] https://salims.pages.iu.edu/gswlc/
[3] https://live-sdss4org-dr14.pantheonsite.io/spectro/galaxy_mpajhu/

Table 1 Overview of sample sizes.

|  | detected | undetected | total |
|---|---|---|---|
| accompanied AGN[a] | 13 | 309 | 322 |
| isolated AGN[b] | 5 | 10 | 15 |
| accompanied non-AGN | 4415 | 10137 | 14553 |
| isolated non-AGN | 1135 | 1799 | 2934 |

[a]including 11 detected and 6 undetected by FAST.
[b]including 5 detected and 2 undetected by FAST.

## 3 FAST OBSERVATIONS AND DATA REDUCTION

### 3.1 FAST observations

We carried out ∼ 16-hour H I line observations on the 27 dwarf galaxies with FAST during 2021 to 2023 (PID: ZD2021_4_2, PI: Wu; PID: PT2022_0182, PI: Chen).

During the observations, we adopted the ON-OFF mode with the 19-beam receiver in L-band (1.05–1.45 GHz, Li et al. 2018). The Spec W+N backend with a 500-MHz-bandwidth is applied to achieve a resolution of 7.63 kHz (1.6 km s$^{-1}$ at 1.4 GHz) per channel over 65,536 channels. Each of the 19 beams has a full width half power (FWHP) of ∼ 2.9′ at ∼ 1.42 GHz. The observations were split into several ON/OFF cycles consists of a 300 or 450 s ON-source and OFF-source integrations with a sampling time of ∼ 0.1 s. The OFF-source position is set to be (−11.8′, 0′) relative to the given ON-source coordinates in R.A and Dec, so that the source can be observed by central beam (M01) and the outer-most beam (M08/M14) of the 19-beam receiver at ON and OFF positions, respectively. This observing strategy was employed in other FAST observations (e.g. Zheng et al. 2020; Yu et al. 2024). For sources with potential H I sources in the default OFF-position based on optical images, we used the offset position of (11.8′, 0′), which corresponding to the position of the west-most beam (M08). During ON-source observations, beam M01 is pointed to the position of the target, while beam M08 or M14 is pointed to the OFF position simultaneously. Correspondingly, during the OFF-source observations, beam M01 is pointed to the selected OFF position with beam M14 or M08 pointing to the source. This ON-OFF observation configuration allowed us most efficiently accumulating the integration time on the targets. The final effective integration time is about 10 to 60 minutes for each source, reaching an rms noise level of ∼ 0.3 to 1 mJy with a chanenel width of 6.4 km s$^{-1}$. We also record high cadence data from the Pulsar backend to trace the volatility of the RFI, providing reference to flag the contaminated data in the spectroscopic backend data (see Section 3.2 for details).

### 3.2 Data reduction

The observed FAST data were processed with self-developed data reduction procedures written in Python 3.8. The system temperature ($T_{sys}$) of all sampled spectra were calibrated using the high cadence noise diode during observations. After that, spectra with abnormal median $T_{sys}$ are removed using a 3-sigma clipping of the global median through iterations to minimize RFI contamination. Then the 0.1-s spectral data were rebinned to 60-s chunks. We mask the frequencies where detected emissions are presented, and subtract the smoothed baseline from the rest of the spectra using a Gaussian kernel with $\sigma$ = 100 channels. To reduce the standing wave of the baseline due to the reflections between the primary mirror and the receiver chamber, each 60-s spectrum was fitted with a sinusoidal component over a range of 5 MHz around the central frequency, with the measured H I line emission masked. We then employed noise-weighted averaging of the data from two polarizations of all the 60-s integrations. The final spectra of our targets were obtained by subtracting the sky spectrum detected from the corresponding OFF positions (M08 or M14). Similar processes were done to the data observed from the other beams (M02 to M19 except M08/M14) that are pointed to the surroundings of the targets. Finally, frequency regions with severe RFI are manually masked before follow-up calculation and analysis. After RFI removal, we converted the measured temperature of the obtained spectra into flux densities using the telescope gain, which is a function of beam, frequency, and zenith angle (Jiang et al. 2020). The final spectra (as shown in Section 4) were rebinned by every 4 adjacent channels and converted to heliocentric velocities, resulting a spectral resolution of 6.4 km s$^{-1}$.

We estimate the signal-to-noise ratio (S/R) in the expression of:

$$S/N = \left(\frac{S_{int}}{W_{50}}\right) \frac{n_{ch}^{1/2}}{\sigma_{rms}}, \quad (1)$$

following Haynes et al. (2018), where $S_{int}$ denotes the integrated flux density; $W_{50}$ represents the line width at 50% of the peak flux; $n_{ch}$ is the number of channels within $W_{50}$ with a channel width of 6.4 km s$^{-1}$. $\sigma_{rms}$ is the standard deviation of the spectra calculated over a velocity range ± 500 km s$^{-1}$ around but excluding the central velocity of the H I profile region. We identify detections for sources with a S/N > 3. For undetected sources, the upper limit to their integral flux densities are estimated by $S_{lim} = 3\sigma_{rms} \times \sqrt{\frac{150}{6.4}}$ km s$^{-1}$, assuming a typical $W_{50}$ of ∼ 150 km s$^{-1}$, which is close to the mean $W_{50}$ of our detected sources. H I mass ($M_{H\,I}$) is derived following Meyer et al. (2017):

$$\frac{M_{H\,I}}{M_\odot} = \frac{2.35 \times 10^5}{(1+z)^2} \left(\frac{d_L}{Mpc}\right)^2 \left(\frac{S_{int}}{Jy\ km\ s^{-1}}\right), \quad (2)$$

where $d_L$ represents the luminosity distance. For the undetected sources, the upper limit to their $M_{H\,I}$ is constrained by the values of $S_{lim}$ above.

## 4 RESULTS

### 4.1 H I detections

Out of the 26 observed galaxies, we had clear detections from 18 targets including 17 sources with AGN and 1 non-AGN SF dwarf galaxy. A summary of the physical and H I properties of the sources are given in Table 2. For the undetected sources, we calculate the 3$\sigma$ upper limits. Example spectra are shown in Figure 1. All spectra for each of the targets are presented in Figure A1 in the **Appendix**.

In addition to H I detections within the beam at the target positions, there are 11/25 AGN-host dwarfs having H I detection in their nearby environment detected by outer beams of FAST 19-beam receiver. These H I detections have a central velocity within 500 km s$^{-1}$ offset to the central sources. The 19-beam spectra around these targets overlaid on SDSS optical images are shown Figure A2 in the **Appendix**. We list properties of the H I emission lines detected with outer beams in Table 3. Information of their possible optical counterparts recorded in SIMBAD or NED are listed in Table 4. We have also searched for possible optical counterparts recorded in the DESI Legacy Survey (Dey et al. 2019), however, no additional sources have been identified. We note that two detected sources

**Table 2** Basic information and H I properties of the targets.

| Source Name | Name in Literature | Obs. Epoch yy/mm/dd | $z^{\rm opt}_{\rm spec}$ | $f_{\rm int}$ [mJy km s$^{-1}$] | $f_{\rm peak}$ [mJy] | r.m.s [mJy] | $v^{\rm H\,I}_{\rm helio}$ [km s$^{-1}$] | $V_{85}$ [km s$^{-1}$] | log $M_{\rm H\,I}/M_\odot$ | log $M_*/M_\odot$ |
|---|---|---|---|---|---|---|---|---|---|---|
| (1) | (2) | (3) | (4) | (5) | (6) | (7) | (8) | (9) | (10) | (11) |
| J021934.93−002432.3 | NSA7119 | 2023/05/23 | 0.026 | 147.9 ± 4.6 | 1.5 | 0.3 | 7719 ± 3 | 118 ± 6 | 8.63 ± 0.13 | 8.9 |
| J043242.28−042257.0 | NSA171487 | 2022/07/28 | 0.015 | 365.8 ± 6.3 | 3.7 | 0.5 | 4292 ± 3 | 186 ± 6 | 8.54 ± 0.13 | 9.4[c] |
| J080028.55+152711.2 | NSA104881 | 2022/09/02 | 0.015 | 1291.7 ± 11.7 | 12.1 | 0.4 | 4533 ± 2 | 258 ± 2 | 9.11 ± 0.13 | 9.1[c] |
| J081145.29+232825.7[a] | J08+23 | 2021/12/26 | 0.016 | 137.7 ± 2.9 | 3.2 | 0.3 | 4715 ± 1 | 63 ± 1 | 8.18 ± 0.13 | 9.2 |
| J084025.54+181858.9[a] | J08+18 | 2022/04/04 | 0.015 | 58.8 ± 1.5 | 0.6 | 0.1 | 4553 ± 5 | 191 ± 3 | 7.75 ± 0.13 | 9.2 |
| J094419.41+095905.3 | NSA51928 | 2022/09/02 | 0.010 | 768.7 ± 7.5 | 4.1 | 0.4 | 3055 ± 3 | 322 ± 2 | 8.54 ± 0.13 | 9.6 |
| J095418.16+471725.1[a] | J09+47 | 2022/04/10 | 0.033 | < 514.0 | 0.6 | 0.2 | – | – | < 7.99 | 9.4 |
| J101440.21+192449.0[b] | J10+19 | 2022/04/10 | 0.029 | 264.6 ± 4.8 | 2.0 | 0.2 | 8648 ± 4 | 276 ± 5 | 8.97 ± 0.13 | 8.9 |
| J114040.73+594850.5 | NSA124554 | 2022/09/04 | 0.012 | < 1234.7 | 1.6 | 0.6 | – | – | < 7.50 | 9.0 |
| J115704.07+221845.5 | NSA113900 | 2022/09/04 | 0.023 | 411.0 ± 4.2 | 3.1 | 0.4 | 6921 ± 2 | 189 ± 3 | 8.97 ± 0.13 | 9.0 |
| J121042.46+131848.4 | NSA66491 | 2022/09/04 | 0.023 | 4118.1 ± 7.4 | 24.3 | 0.4 | 6924 ± 1 | 396 ± 1 | 9.96 ± 0.13 | 9.1 |
| J121639.46+141537.0 | NSA75400 | 2023/04/28 | 0.023 | 1800.6 ± 5.6 | 13.1 | 0.2 | 7077 ± 1 | 188 ± 1 | 9.62 ± 0.13 | 8.4 |
| J122647.95+074017.6 | NSA67333 | 2022/07/20 | 0.002 | < 1097.0 | 1.1 | 0.5 | – | – | < 5.91 | 8.3[c] |
| J122904.33+423439.5 | NSA61000 | 2022/09/04 | 0.026 | 366.4 ± 5.0 | 2.7 | 0.4 | 7704 ± 3 | 219 ± 3 | 9.01 ± 0.13 | 9.1 |
| J123702.27+065531.0 | NSA141727 | 2022/07/21 | 0.005 | 6200.3 ± 8.5 | 76.8 | 0.4 | 1685 ± 0 | 135 ± 0 | 8.89 ± 0.13 | 9.0[c] |
| J125756.76+275930.6 | NSA103730 | 2023/04/28 | 0.015 | 205.0 ± 5.8 | 1.5 | 0.2 | 4641 ± 4 | 163 ± 7 | 8.31 ± 0.13 | 8.1 |
| J125815.27+272752.9 | NSA142638 | 2022/09/04 | 0.025 | < 991.9 | 1.2 | 0.5 | – | – | < 8.08 | 9.1[c] |
| J132941.93+103636.8 | NSA78105 | 2023/04/28 | 0.027 | 821.0 ± 5.1 | 6.0 | 0.2 | 8029 ± 1 | 163 ± 1 | 9.40 ± 0.13 | 8.7 |
| J153256.96+462700.8 | NSA166259 | 2022/09/02 | 0.002 | 3027.9 ± 9.5 | 67.8 | 0.4 | 674 ± 0 | 99 ± 0 | 7.80 ± 0.13 | 7.8[c] |
| J155206.94+410105.4 | NSA43531 | 2023/05/23 | 0.025 | 990.6 ± 5.6 | 8.4 | 0.3 | 7606 ± 1 | 139 ± 0 | 9.43 ± 0.13 | 8.7 |
| J161902.49+291022.2 | NSA57867 | 2023/05/23 | 0.008 | 522.8 ± 2.9 | 5.0 | 0.2 | 2542 ± 0 | 100 ± 1 | 8.21 ± 0.13 | 7.8[c] |
| J162758.43+390704.3 | NSA46318 | 2022/09/02 | 0.028 | < 1164.8 | 1.4 | 0.5 | – | – | < 8.23 | 9.1 |
| J162812.10+404719.0 | NSA29496 | 2022/09/02 | 0.029 | < 930.6 | 1.3 | 0.4 | – | – | < 8.16 | 9.1 |
| J162843.38+403219.0 | NSA29476 | 2022/09/02 | 0.028 | < 1030.6 | 1.3 | 0.5 | – | – | < 8.18 | 9.1 |
| J162847.61+393859.1 | NSA166937 | 2022/09/02 | 0.027 | 236.8 ± 4.9 | 2.6 | 0.4 | 8245 ± 5 | 214 ± 3 | 8.87 ± 0.13 | 9.2[c] |
| J164238.46+223309.1 | NSA167047 | 2023/05/24 | 0.017 | < 408.5 | 0.6 | 0.2 | – | – | < 6.96 | 8.3 |

(1) Name of the sources, (2) name of the sources in Manzano-King et al. (2019) or in Baldassare et al. (2020a), (3) date of observation, (4) spectroscopic redshift from SDSS, (5) integrated flux density, (6) peak flux density, (7) r.m.s level of the spectrum, (8) heliocentric velocity measured as the midpoint between the channels at which the line flux density is 50% of its peak, (9) the velocity width that captures 85% of the total flux, which most effectively represents the rotation velocity of the galaxy (Yu et al. 2020), (10) measured H I mass using Equation 3.2. For the undetected sources, a 3-$\sigma$ upper limit assuming $W_{50}$ = 150 km s$^{-1}$ is given, assuming 10% uncertainty in the sources distance and a 15% uncertainty in flux, the typical uncertainty of $M_{\rm H\,I}$ is 0.13 dex, (11) stellar mass given in the NSA or GSWLC catalogue. The default source of stellar mass information is the GSWLC catalogue, those from the NSA are labelled.
[a] AGN or composite dwarf galaxy with optical outflow in Manzano-King et al. (2019).
[b] Star-forming dwarf galaxy with optical outflow in Manzano-King et al. (2019).
[c] Stellar mass from K-correction fit for Sersic fluxes provided by NSA.

(J101440.21+192449.0 and J153256.96+462700.8) show H I central velocities more than 300 km s$^{-1}$ with respect to their optical velocities. The H I detection may come from nearby environment or galaxies within the ~ 2.9′ FAST beam. In the statistical analysis in the following sections, we do not include these two sources. We discuss possible origin of the H I signals from the nominated positions of J101440.21+192449.0 and J153256.96+462700.8 in Section 5.1.

tivity limit, we use their corresponding upper limits based on the $\alpha$.100 sensitivity to do the survival analysis. The survival function $S(t)$ built by Kaplan–Meier estimator is inverse to $S(-t)$, i.e, $S(t) = -S(-t)$ (Feigelson & Nelson 1985) where t is the "survival time". In our analysis, we replace t to H I mass ratio $f_{\rm H\,I}$. We performed log-rank tests between each of the subsamples to test if they share similar distributions in $f_{\rm H\,I}$ as shown in 2.

### 4.2 H I mass distributions

We compare the H I mass fraction distributions of the four subsamples (isolated non-AGN, accompanied non-AGN, isolated AGN, and accompanied AGN) by survival analysis with the null hypothesis of no significant difference seen between two samples. To prevent biases from mis-matched sensitivities of FAST and $\alpha$.100, for the FAST observed targets, we estimated their upper limits in $\alpha$.100 based on its 50% completeness (Haynes et al. 2011). For FAST observed sources with an $M_{\rm H\,I}$ or upper limits of $M_{\rm H\,I}$ exceeding the $\alpha$.100 sensi-

In the result of the survival analysis, due to the limited sample size, survival analysis between isolated dwarfs generate a wide 95% confidence interval (from < 0.01 to 0.93; Figure 2a & c). Therefore, no definitive conclusion can be made from the comparisons between gas abundance of AGN-hosting dwarfs and the non-AGN counterparts. On the other hand, as illustrated by Figure 2b & d, in accompanied galaxies, AGN-hosting dwarfs appear to have distinct Kaplan–Meier curves comparing to their non-AGN counterparts with $p < 0.05$.

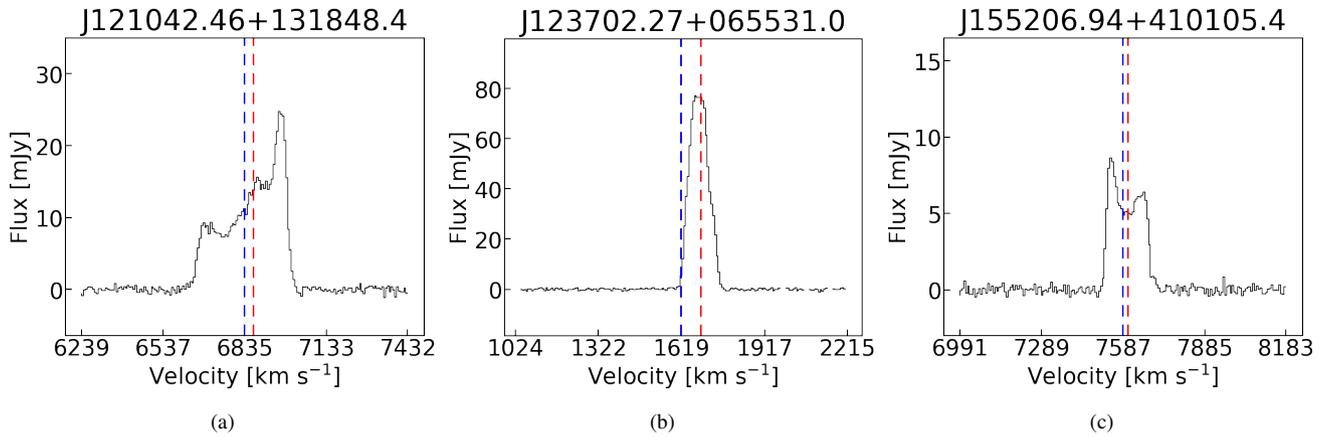

**Figure 1.** Example H I spectrum of the targets. The spectrum are stacked over the central beam (Beam 01) and the outer beam (Beam 08 or 14). The blue lines label the optical velocity calculated from redshift. For the detected sources, the red vertical lines denote the central heliocentric velocity of the H I signal. The H I spectra for all targets are displayed in A1.

### 4.3 SFR distributions

To compare the SFR distributions of dwarf galaxies in the four samples, we plot the histogram of their GSWLC-X2 SFR in Figure 3. The accompanied sources have a wider SFR distribution extending below $\sim 10^{-2}$ $M_\odot$ yr$^{-1}$, while there are only a few isolated dwarfs with SFR lower than that. We performed 1000 times boot-strapping Kolmogorov-Smirnov (K-S) tests between the subsamples. The 95% confidence interval and median of the $p$-values are labeled on the sub-panels of Figure 3. For the non-AGN sources, a small $p$-value of $< 0.05$ indicates distinct SFR distributions from Figure 3a. On the other hand, though the median values of $p$ for the K-S tests between isolated and accompanied AGN (Figure 3b) are lower than 0.05, the broad span of the 95% confidence interval reflects large uncertainties of the test results, therefore, no statistically significant conclusion can be drawn.

Our comparisons on SFR of four subsamples show that the SFR of the dwarf galaxies are highly affected by the environmental conditions, mostly for the systems in the low SFR range. The AGN existence in the dwarf systems shows little impact to the SFR distributions. Previous studies have shown that massive galaxies which are the central galaxy in their dark matter halo are predominantly quenched: the quenched fractions of isolated galaxies is 100% for galaxies with stellar mass above $10^{11}$ $M_\odot$, decreasing to 20% of galaxies with stellar mass of $10^{10}$ $M_\odot$ (Wetzel et al. 2012; Peng et al. 2012; Woo et al. 2013). Geha et al. (2012) found a threshold of $10^9$ $M_\odot$ below which quenched galaxies are barely found in the local Universe. Our result in Figure 3 is consistent to Geha et al. (2012) that there is a lack of low-mass dwarf galaxies with low SFR in isolated galaxies.

## 5 DISCUSSIONS

### 5.1 AGN feedback in dwarf galaxies

The role of AGN feedback in dwarf galaxy evolution has long been debated (see Manzano-King et al. 2019 for a review). The potential of AGN feedback suppressing star formation in dwarf galaxies has been demonstrated from multiple simulation works (e.g. Barai & de Gouveia Dal Pino 2019; Koudmani et al. 2022; Arjona-Galvez et al. 2024). Dashyan et al. (2018) argued that AGN could potentially have played a significant role in gas ejection in early dwarf evolution. In a series of papers, Zhang et al. (2020); Cai et al. (2020, 2021) show that their 60 AGN-host dwarfs galaxies in their samples are 3–7 times older than that without an AGN. They also observed a mild correlation between the star formation histories (SFHs) and AGN activity indicating that they may be fueling the same gas reservoir. Penny et al. (2018) demonstrate that AGN feedback may keep some bright dwarf elliptical (dE) galaxies quiescent in galaxy groups and the outskirts of clusters.

However, the impact of AGN in the low-mass Universe remain controversy. Shangguan et al. (2018) show that the AGN activities have minor effect of the host SFRs. Koudmani et al. (2019) suggest that AGN activity can be surpassed by supernovae (SNe) and only affect the global SF indirectly by enhancing SN feedback. Trebitsch et al. (2018) also show that AGN feedback is negligible compared to stellar activity in such systems. Besides, Ward et al. (2022) argue that the presence of AGN feedback does not appear to coincident with lower instantaneous SFRs during individual star formation episodes.

To assess the AGN activity impacts on the H I contents in the low-mass galaxies environment, we examine the $f_{\rm H\,I}$ distributions of AGN and non-AGN-host dwarfs in isolated environments. We plot the $f_{\rm H\,I}$ distributions of isolated AGN in comparison to their non-AGN counterparts in Figure 2a. The KM functions of $f_{\rm H\,I}$ in isolated galaxies are displayed in Figure 2c. The black dotted line denotes the 50% survival probability of $f_{\rm H\,I}$, at which the median values of $f_{\rm H\,I}$ are estimated. As a result, in isolated systems, dwarf galaxies hosting an AGN appear to have lower median $f_{\rm H\,I}$, which may indicate their relatively gas-poor nature. However, due to a limited sample size, the 1000-time bootstrapped log-rank test exhibits a large spread. The 95% confidence interval of the $p$-value ranges from $< 0.01$ to 0.81. Consequently, within isolated dwarfs, no definitive conclusions can be drawn regarding whether the presence of an AGN are associated with a reduced $f_{\rm H\,I}$ comparing to the control sample.

On the other hand, for accompanied dwarfs, as illustrated by Figure 2b & d, AGN-hosts show distinctively lower $f_{\rm H\,I}$ to the non-AGN dwarf galaxies with $p_{97.5} < 0.05$, indicating their relatively gas-deficient nature. This may be supportive to the scenario in which AGN could effectively reduce the cold gas abundance in dwarf galaxies. Indeed, low-luminosity AGN may be responsible for heating ambient gas and thereby suppress star formation in the host galaxies (Cheung et al. 2016; Wang et al. 2024). Similar observational evidence is also found in dwarf galaxies. As suggested by Penny et al.

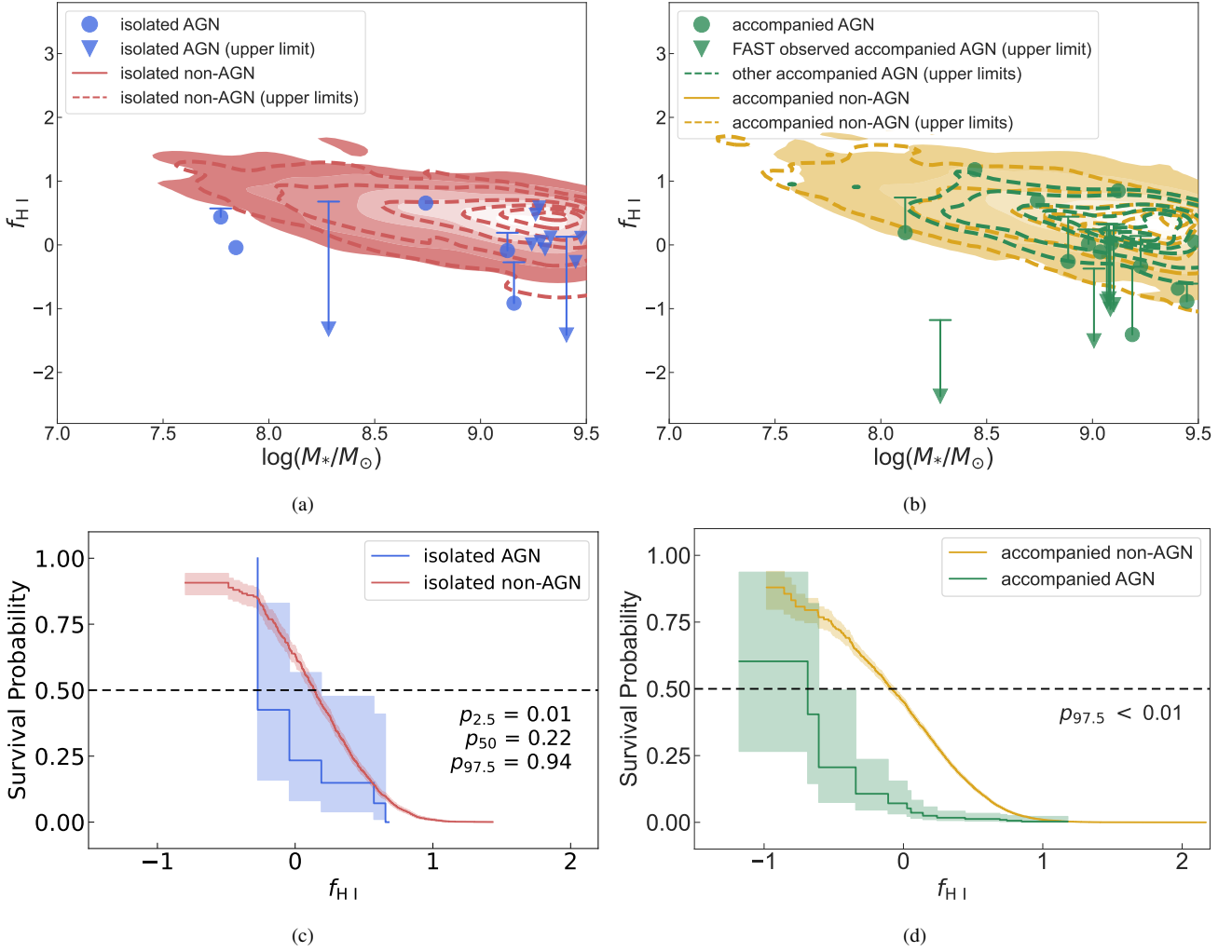

**Figure 2.** Distributions of $f_{\mathrm{H\,I}}$ along $M_*$ of (a) isolated and (b) accompanied dwarf galaxies. The filled dots represent the AGN-hosts including FAST observed targets and those in the footprint. The downward triangles represent the $3\sigma$ upper limits to the undetected sources. For FAST observed targets, if the measured flux or estimated upper limit is deeper than the $\alpha$.100 sensitivity threshold, we mark the corresponding $\alpha$.100 detection limit estimated at its 50% completeness by a short horizontal bar, connected to their FAST measured data by a vertical line. The filled contours locate the $\alpha$.100 detected non-AGN dwarf galaxies. $3\sigma$ upper limits to undetected sources are illustrated by the dotted contour curves under the assumption that $W_{50} = 150$ km s$^{-1}$. The contour lines are plotted at the 0%, 5%, 25%, 50%, 75%, and 95% peak source densities. Kaplan-Meier survival curves of $f_{\mathrm{H\,I}}$ of (c) isolated and (d) accompanied AGN (FAST + $\alpha$.100 undetected) and non-AGN ($\alpha$.100 detected + undetected) samples. For the FAST observations that beyond the $\alpha$.100 50% completeness sensitivity, we use their matched $\alpha$.100 sensitivity to build the survival function. The 2.5%, 50%, and 97.5% percentiles of $p$-value from the 1000-time log-rank test are displayed on the right-hand side of the panels. Notably, the KM curves are plotted over the full samples rather than the boot-strapped samples. The dotted horizontal line labels their median at 50% survival probability.

(2018), although environmental effects are still likely the dominant star formation quenching mechanism for dE galaxies, AGN feedback may prevent gas cooling in low-mass galaxies, and consequently help some dEs maintain quiescence after the initial quenching episode, i.e. the so-called "maintenance-mode" AGN feedback.

### 5.2 Environmental effects on dwarf galaxies

Since environmental effects can efficiently impact the gas reservoirs of galaxies, it can therefore potentially affect the fuelling of their central black hole as well. For example, a lower number density of AGN is observed in isolated systems potentially due to a lack of gas supply to feed the SMBH (Sabater et al. 2013), or less frequent interactions between galaxies that could trigger AGN activity, which often occur in denser regions (Gisler 1978; Dressler et al. 1985; Popesso & Biviano 2006; Pimbblet et al. 2013; Lopes et al. 2017).

In contrast, some other studies reported a low dependency of AGN fraction on densities (Miller et al. 2003; Martini et al. 2006; Rodríguez del Pino et al. 2017). In particular, Amiri et al. (2019) found that AGN-host galaxies have similar distributions of specific star formation rates and of galactic ages both in clusters and in voids. Kauffmann et al. (2004) found that AGN host galaxies of strong [O III]-emission as an indication of higher AGN activity are twice as frequent in low density regions as in high density regions. Montero-Dorta et al. (2009) also find a slowly declining AGN fraction towards high-density environments in the SDSS. Moreover, Rodríguez Del Pino et al. (2023) observed a negative connection between density and AGN activity, suggesting that AGN in dense environments accrete less gas than those isolated. However, the declination of AGN-driven outflow towards denser region is only significant within lower-mass galaxies ($10^9 - 10^{10.3} M_\odot$). The suppression of AGN outflow may indirectly induce an higher metallicity seen in clustered galaxies, especially at lower masses.

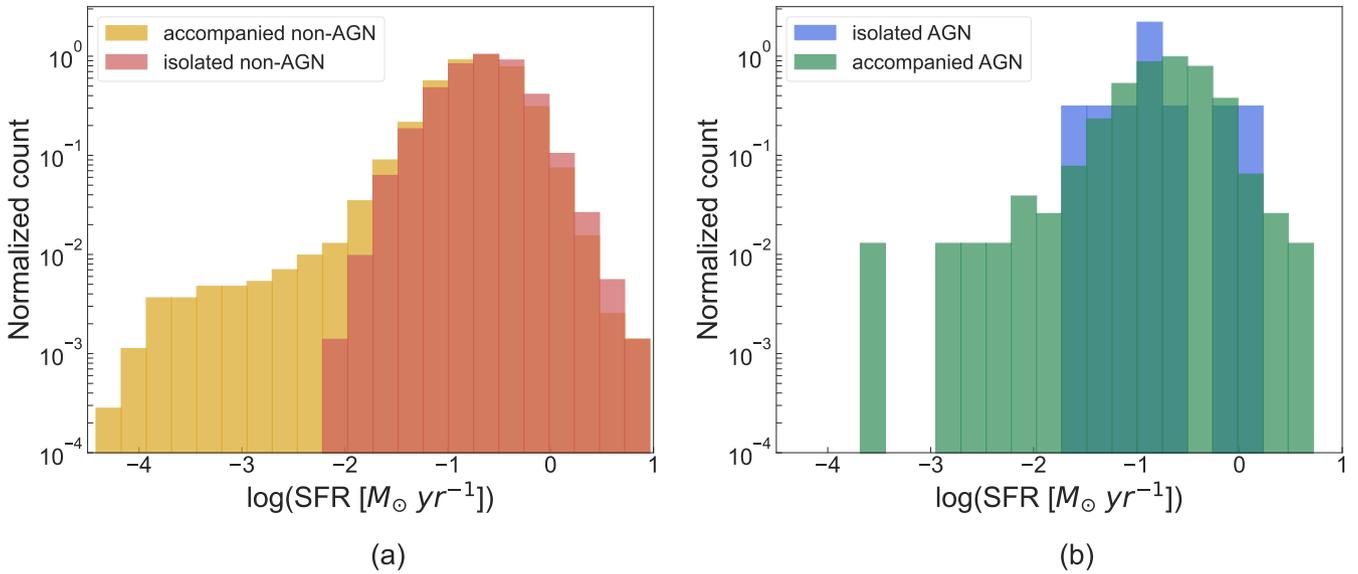

**Figure 3.** GSWLC SFR histograms of (a) isolated and accompanied non-AGN, (b) isolated and accompanied AGN.

One possible mechanism through which the environment may alter the AGN activity is merging. Merging events frequently happen in the field, and they are widely believed to be triggers of AGN activity (e.g. Sanders et al. 1988; Bahcall et al. 1997; Canalizo & Stockton 2001; Urrutia et al. 2008; Letawe et al. 2010; Smirnova et al. 2010. Another process capable of affecting the gas supply in galaxies is ram pressure stripping (Gunn & Gott 1972b), which happens most frequently in clusters and massive groups (Hester 2006), or in lower-density groups with halo masses of approximately $10^{13} M_\odot$ (Roberts et al. 2021; Kolcu et al. 2022). Ram pressure stripping can effectively remove the gas content away from the ISM of galaxy (van Gorkom 2004; Kenney et al. 2004; Poggianti et al. 2017; Fumagalli et al. 2014) and inducing a quenching of star formation (Vollmer et al. 2001; Tonnesen et al. 2007; Vulcani et al. 2020). However, it has been proposed that ram pressure can initially boost star formation by gas compression before gas removal both theoretically (Kronberger et al. 2008; Kapferer et al. 2009; Tonnesen & Bryan 2009; Bekki 2014), and observationally (Crowl & Kenney 2006; Merluzzi et al. 2013; Vulcani et al. 2018; Peluso et al. 2022). The same process may also ignite the AGN by gas inflow caused by gravitational loss of angular momentum (Schulz & Struck 2001; Tonnesen & Bryan 2009; Ramos-Martínez et al. 2018). Conversely, Joshi et al. (2020) suggested that the enhancement of BH mass accretion is significantly suppressed in clusters for galaxies with $M_* \sim 10^{9.7-11.6} M_\odot$.

On the other hand, the environment-AGN connection seems to be mass-dependent. While ram-pressure stripping may trigger AGN activity in galaxies with $M_* > 10^{9.5} M_\odot$, less massive systems may be suppressed in both star formation and black hole accretion (Marshall et al. 2018; Ricarte et al. 2020). Indeed, Peluso et al. (2022) find an AGN fraction of $\sim 27\%$ in for ram pressure stripped galaxies with stellar masses above $10^9 M_\odot$. This fraction raises to 51% for galaxies with $M_* > 10^{10} M_\odot$. Rodríguez Del Pino et al. (2023) claimed that the presence of ionized outflows increases strongly towards higher masses (from $\sim 7\%$ to 32%) in accompanied AGNs within a mass range of $\sim 10^{9.0-11.5} M_\odot$. However, no significant variation within the same mass range is found in the isolated sample. In addition, the selection methods of AGN also seem to affect the observed influence from the environment (e.g. Roman-Oliveira et al. 2019; Peluso et al. 2022; Rodríguez Del Pino et al. 2023).

Our observational results provide evidences to support the elevated AGN activity in dense environment. First of all, in our study, $\sim$ 2.2% (322/14847) galaxies are identified to be a host of an AGN in accompanied dwarfs, which is nearly 4 times to the $\sim 0.6\%$ (17/2951) AGN fraction within isolated sources. Secondly, as mentioned in Section 4.2 and 5.1, we observed a lower $f_{\rm H\,I}$ for accompanied AGN in comparison to the non-AGN control sample (Figure 2d). In fact, dense IGM may have positive an interplay with AGN feedback in early stages of interaction, i.e the "two-phase environmental effects". For example, some studies have claimed that SFR and BH accretion may be initially enhanced by gas compression before a complete gas removal in the later stage(Kronberger et al. 2008; Kapferer et al. 2009; Tonnesen & Bryan 2009; Bekki 2014; Crowl & Kenney 2006; Merluzzi et al. 2013; Vulcani et al. 2018; Peluso et al. 2022).

On the other hand, environmental effect may effectively hamper SFR in dwarf galaxies: Firstly, as indicated by Figure 4, SFR for both AGN and non-AGN samples slightly increase along larger distance to nearest massive galaxies. Their positive scaling relation with Pearson's coefficients of 0.21 and 0.16 are significant with $p-$values less than 0.05, indicating a suppressed SFR towards denser environments. Moreover, there are barely quiescent galaxies among our isolated dwarfs (see Section 4.3 and Figure 3), which is in good agreement to previous findings (Geha et al. 2012). This may indicate that environmental effect is still the major cause of quenching in low mass galaxies. Due to a lack of hot halo, it is theoretically predicted that stellar feedback in dwarf galaxies can temporarily quench star formation, but the ejected gas will cool down and be recycled for to fuel new episodes of star formation (e.g. Dekel & Silk 1986; Stinson et al. 2007; Bradford et al. 2015). It is reasonable to assume that similar condition may also happen in AGN-driven gas ejection. This could in turn cause a low incidence of quiescent isolated dwarfs.

In conclusion, although there are evidences indicating a depleted neutral gas seen in dwarfs galaxies with an AGN, the major factor

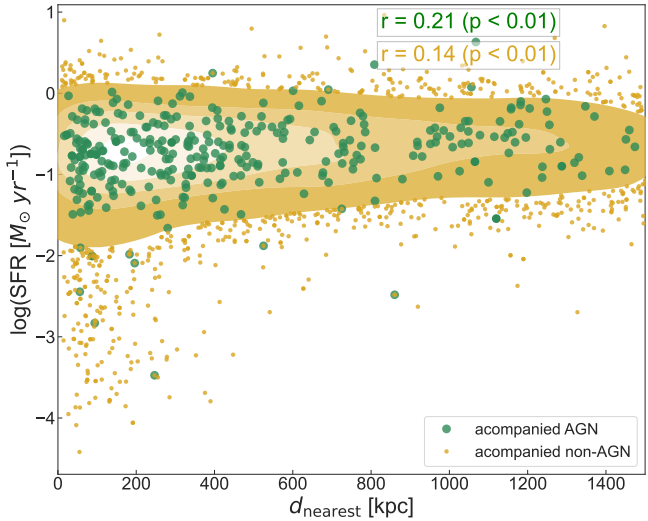

**Figure 4.** SFR distributions of AGN and non-AGN accompanied samples along projected distance to the nearest massive galaxies. The green points displays the isolated AGN-hosting dwarfs. The orange contours represent the 5%, 25%, 50%, 75%, and 95% peak source densities of isolated non-AGN dwarf galaxies. The correlation coefficients and their $p$−value for the two sub-samples is presented on the upper right corner of the panel.

that are responsible of quenching dwarf galaxies is likely from the external environments.

### 5.3 Possible origins of the H I emissions from near AGN-host dwarf galaxies

The 19-beam receiver of FAST in L band allows us to detect possible H I gas in the surrounding environment of the galaxies (∼ 50−200 kpc at the redshifts of the targets; See Figure A2). In our observational campaign, 11 out of 27 dwarf galaxies exhibit H I detections in the FAST side beams, they are all accompanied AGN. In addition, we also detected a H I source (beam 18) near J043242.28042257.0 with a large radio velocity offset of 683 km s$^{-1}$. In total, there are 18 H I emission lines detected in the surroundings of 11 sources. We search for possible associated galaxies of these 20 H I sources from NED and SIMBAD databases with a search radius of 3′ (FWHP of FAST), and an optical velocity offset of < 500 km s$^{-1}$ to the detected H I line. The measured properties of the H I emission lines detected by the outer beams are listed in Table 3. The information of their possible optical identifiers as well as stellar mass information from the NMJG catalogue are included in Table 4.

We find that 13 H I sources near 9 targeted dwarf galaxies have been attributed to known galaxies nearby. In particular, J043242.28042257.0 have detected H I emission in 5 outer beams at projected distances of ∼ 100−300 kpc, reflecting a very dense environment. This source possibly resides in a galaxy group where ram pressure stripping becomes prominent. Besides, 5 emission line sources near 5 targeted dwarf galaxies are not associated with any known identifiers (see Table 4). They locate at ∼ 280−400 kpc away from the central sources, largely beyond the gravitation bound of the central dwarf galaxies. These emissions are possible tracers of environmental effects such as ram pressure stripping (van Gorkom 2004; Kenney et al. 2004; Poggianti et al. 2017; Fumagalli et al. 2014). Another possible origin to these H I clouds is gas-dominated baryonic systems in sub halos within a larger halo which also covers the targeted dwarfs.

### 5.4 Asymmetric H I emissions from AGN-host dwarfs

As presented in Figure 1, we found several sources show asymmetric non-Gaussian profile in their spectra (J084025.54+181858.9, J121042.46+131848.4, J121639.46+141537.0, J132941.93+103636.8, J155206.94+410105.4 and J161902.49+291022.2). On the other hand, J121042.46+131848.4 and J132941.93+103636.8 have three peaks in their H I line profiles. We also note that the velocity dispersions we measured in this paper are not inclination-corrected. Therefore we anticipate our measurements are lower limits to the actual velocities. Some systems may have inflow/outflow. However, no explicit features are observed in our data.

Aside from stellar or AGN feedback, the kinematical perturbation induced asymmetric H I profile may also arise from other external factors such as environmental effects (e.g., Cortese et al. 2021 and references therein), merger events (e.g., Jog & Solomon 1992; Barnes 2002; Boomsma et al. 2005; Robertson et al. 2006), or gas accretion (Bournaud et al. 2005; Sancisi et al. 2008). In addition, some internal processes may also generate similar features as well. For example, warp in the disc arise from accretion of gas in the intergalactic median (CGM) (López-Corredoira et al. 2002), a misaligned halo (Ostriker & Binney 1989) or an intergalactic magnetic field (Battaner et al. 1990), or non-axisymmetric gravitational potential (Baldwin et al. 1980; Hayashi & Navarro 2006) can be associated with the presence of a bar (Saha et al. 2007; Newnham et al. 2020) or spiral arms (Laine & Gottesman 1998).

The sources mentioned above with asymmetric H I profiles are ideal systems for studying environmental and AGN effect on low-mass galaxies. To confirm the origin of their H I emissions, high angular resolution observations on their H I distribution are required.

### 5.5 Caveats of this study

Due to significant better sensitivities from longer integrations, our FAST observation can detect less gas-rich galaxies in our sample comparing to the $\alpha$.100 catalogue. Besides, while our sample of AGN with optical outflow or MIR variability is selected from the literature, the ionized gas outflows and mid-IR variabilities of galaxies in the control sample is not carefully examined accordingly. This may potentially induce discrepancies of physical properties between the FAST observed sample and the control sample. In addition to the possible selection effect, it is also notable that some of the dwarf galaxies in the control sample might be contaminated by weak AGN. For example in multi-band studies, AGN are estimated to occupy the dwarf population at the level of < 1% (Sartori et al. 2015; Reines et al. 2013; Moran et al. 2014). But yet, a more recent study reveals an AGN fraction of ∼ 20% within dwarf galaxies z < 0.15 (Mezcua & Domínguez Sánchez 2024). Furthermore, this study has a limited sample size of AGN-dwarfs. Future large-scale studies on the impacts of AGN activities on the ISM gas in and around the dwarf galaxies can provide stronger constraints to evaluate the conclusions in this work.

## 6 SUMMARY

In this paper, we report H I measurements and analysis on samples of dwarf galaxies with or without identified AGN features using FAST observations and the ALFALFA survey. We further divide them into isolated and accompanied subsamples using optical data.

**Table 3** Information of nearby H I signals detected by FAST outer beams.

| Source Name | RA | DEC | $z_{\rm spec}$ | $f_{\rm int}$ | $f_{\rm peak}$ | r.m.s | $V_{\rm helio}$ | $V_{85}$ | log $M_{\rm H\,I}/M_\odot$ |
| Obs. Beam | hh:mm:ss | dd:mm:ss | | [mJy km s$^{-1}$] | [mJy] | [mJy] | [km s$^{-1}$] | [km s$^{-1}$] | |
| (1) | (2) | (3) | (4) | (5) | (6) | (7) | (8) | (9) | (10) |
| J021934.93−002432.3 | 02:20:20.93 | −00:24:31.6 | 0.026 | 147.9 ± 4.6 | 1.5 | 0.3 | 7719 ± 2.5 | 118 ± 6 | 8.63 ± 0.13 |
| Beam 17 ON | 02:19:34.86 | −00:14:35.9 | | 1882.0 ± 6.2 | 15.1 | 0.5 | 7709 ± 1.4 | 289 ± 1 | 9.73 ± 0.13 |
| J043242.28−042257.0 | 04:33:28.41 | −04:22:56.3 | 0.015 | 365.8 ± 6.3 | 3.7 | 0.5 | 4292 ± 3.3 | 186 ± 6 | 8.54 ± 0.13 |
| Beam 13 ON | 04:32:07.78 | −04:27:56.4 | | 824.3 ± 11.0 | 4.9 | 0.7 | 4253 ± 3.1 | 292 ± 3 | 8.89 ± 0.13 |
| Beam 07 ON | 04:32:53.75 | −04:17:58.2 | | 116.0 ± 9.3 | 2.4 | 0.9 | 4467 ± 6.7 | 89 ± 9 | 8.04 ± 0.13 |
| Beam 12 OFF | 04:31:31.88 | −04:32:54.6 | | 626.4 ± 9.0 | 13.0 | 1.0 | 4030 ± 1.6 | 90 ± 2 | 8.77 ± 0.13 |
| Beam 18 ON | 04:33:05.26 | −04:13:00.0 | | 608.3 ± 11.2 | 9.8 | 1.0 | 4976 ± 2.7 | 103 ± 4 | 8.76 ± 0.13 |
| Beam 19 ON | 04:33:16.82 | −04:17:58.2 | | 1163.1 ± 18.4 | 6.2 | 0.9 | 4456 ± 4.8 | 304 ± 9 | 9.04 ± 0.13 |
| J080028.55+152711.2 | 08:01:16.28 | 15:27:11.9 | 0.015 | 1291.7 ± 11.7 | 12.1 | 0.4 | 4533 ± 1.7 | 258 ± 2 | 9.11 ± 0.13 |
| Beam 09 ON | 08:01:04.27 | 15:22:13.6 | | 743.7 ± 15.3 | 7.9 | 0.6 | 4775 ± 3.3 | 220 ± 4 | 8.87 ± 0.13 |
| J114040.73+594850.5 | 11:42:12.22 | 59:48:51.2 | 0.012 | < 1234.7 | 1.6 | 0.6 | – | – | < 7.50 |
| Beam 07 ON | 11:41:03.54 | 59:53:49.3 | | 2320.4 ± 20.8 | 19.5 | 0.8 | 3668 ± 3.6 | 310 ± 7 | 9.14 ± 0.13 |
| J115704.07+221845.5 | 11:57:53.79 | 22:18:46.2 | 0.023 | 411.0 ± 4.2 | 3.1 | 0.4 | 6921 ± 2.5 | 189 ± 3 | 8.97 ± 0.13 |
| Beam 11 ON | 11:57:04.15 | 22:08:49.1 | | 155.5 ± 7.2 | 2.7 | 0.7 | 6834 ± 5.9 | 135 ± 9 | 8.55 ± 0.13 |
| Beam 16 ON | 11:56:39.09 | 22:28:41.3 | | 238.9 ± 5.2 | 3.2 | 0.5 | 6578 ± 2.8 | 147 ± 4 | 8.73 ± 0.13 |
| J121042.46+131848.4 | 12:11:29.73 | 13:18:49.1 | 0.023 | 4118.1 ± 7.4 | 24.3 | 0.4 | 6924 ± 0.8 | 396 ± 1 | 9.96 ± 0.13 |
| Beam 12 ON | 12:10:19.01 | 13:08:50.8 | | 83.5 ± 3.2 | 1.7 | 0.3 | 6962 ± 2.2 | 74 ± 2 | 8.27 ± 0.13 |
| J125815.27+272752.9 | 12:59:07.11 | 27:27:53.6 | 0.025 | < 991.9 | 1.2 | 0.5 | – | – | < 8.08 |
| Beam 11 ON | 12:58:15.35 | 27:17:56.5 | | 138.6 ± 7.1 | 2.1 | 0.7 | 7488 ± 4.7 | 79 ± 4 | 8.58 ± 0.13 |
| J155206.94+410105.4 | 15:53:07.91 | 41:01:06.1 | 0.025 | 990.6 ± 5.6 | 8.4 | 0.3 | 7606 ± 0.7 | 139 ± 0 | 9.43 ± 0.13 |
| Beam 10 ON | 15:52:37.35 | 40:51:09.6 | | 234.5 ± 5.2 | 6.0 | 0.4 | 7551 ± 1.2 | 47 ± 2 | 8.81 ± 0.13 |
| Beam 11 ON | 15:52:07.04 | 40:51:09.0 | | 80.9 ± 6.0 | 1.8 | 0.4 | 7629 ± 4.6 | 57 ± 7 | 8.34 ± 0.13 |
| J162758.43+390704.3 | 16:28:57.72 | 39:07:05.0 | 0.028 | < 1164.8 | 1.4 | 0.5 | – | – | < 8.23 |
| Beam 12 ON | 16:27:29.06 | 38:57:06.7 | | 304.2 ± 5.7 | 4.5 | 0.6 | 8350 ± 2.6 | 145 ± 2 | 9.01 ± 0.13 |
| J162812.10+404719.0 | 16:29:12.86 | 40:47:19.7 | 0.029 | < 930.6 | 1.3 | 0.4 | – | – | < 8.16 |
| Beam 12 ON | 16:27:42.01 | 40:37:21.4 | | 56.6 ± 6.2 | 2.8 | 0.5 | 8990 ± 2.0 | 21 ± 3 | 8.30 ± 0.13 |
| Beam 17 ON | 16:28:12.00 | 40:57:15.4 | | 155.1 ± 8.3 | 2.1 | 0.7 | 8741 ± 7.7 | 163 ± 7 | 8.74 ± 0.13 |
| J162843.38+403219.0 | 16:29:43.91 | 40:32:19.7 | 0.028 | < 1030.6 | 1.3 | 0.5 | – | – | < 8.18 |
| Beam 13 ON | 16:27:58.17 | 40:27:19.6 | | 157.3 ± 4.3 | 3.6 | 0.5 | 8856 ± 1.5 | 54 ± 1 | 8.72 ± 0.13 |

H I measurements from individual observing sessions are split by lines. (1) Name of the targeted dwarf galaxies and the FAST beam used for HI measurements. The "ON/OFF" tag labels the ON/OFF-source integration when the signal was detected. See Section 3.1 for more details about the observing mode. (2) & (3) central position of the beam, (4) spectroscopic redshift of the central source from SDSS, (5) integrated flux density, (6) peak flux density, (7) r.m.s level of the spectrum, (8) heliocentric velocity measured as the midpoint between the channels at which the line flux density is 50% of its peak, (9) the velocity width that captures 85% of the total flux, which most effectively represents the rotation velocity of the galaxy (Yu et al. 2020), (10) measured H I mass using Equation 3.2. For the undetected sources, a 3-$\sigma$ upper limit assuming $W_{50}$ = 150 km s$^{-1}$ is given. Assuming 10% uncertainty in the sources distance and a 15% uncertainty in flux, the typical uncertainty of $M_{\rm H\,I}$ is 0.13 dex.

We compare the H I abundance and SFR of dwarf galaxies in our sample of AGN vs. non-AGN hosts, or isolated and accompanied systems utilizing their observed H I profile and optical properties. The main conclusions are listed below:

1. AGN are more commonly existed in accompanied dwarf galaxies than in isolated dwarfs (∼ 2.2% and 0.6% AGN fraction, respectively), which may suggest an enhanced black hole accretion caused by gas suppression of nearby galaxies.

2. In accompanied galaxies, AGN-hosting dwarfs have slightly but significantly lower H I mass fraction relative to the non-AGN sample. This may indicate that AGN could regulate the ISM in their low-mass hosts.

3. Accompanied dwarf galaxies appear to have higher SFR towards larger distance to the nearest massive galaxy. Besides, there are only a few isolated galaxies have a SFR of lower than $10^{-2}$ $M_\odot$ yr$^{-1}$. These results suggest that environmental effect may effectively suppress the star formation in dwarf galaxies. It is likely the major quenching mechanism in dwarf galaxies.

4. Utilizing the wide angular coverage of FAST 19 beam receiver in L band, we detected 29 H I line emission sources at ∼ 80 to ∼ 400 kpc away from 10 targeted AGN-host and 1 star-forming dwarf galaxies. We attribute them to be companion gas clouds, or ejected gas by ongoing outflow or environmental processes such as ram pressure stripping. They are worthwhile for case studies with high angular resolution observations.

**ACKNOWLEDGEMENT**

This work is supported by National Key R&D Program of China No. 2023YFE0110500, and National Science Foundation (NSFC) of China No. 12588202 and 12041302. Hongying Chen is supported by the project funded by China Postdoctoral Science Foun-

**Table 4** Information of nearby H I signals detected by FAST outer beams.

| Source Name | Identifier | RA hh:mm:ss | DEC dd:mm:ss | ΔRA ['] | ΔDEC ['] | z | d [kpc] | $\Delta V_{helio}$ [km s$^{-1}$] |
|---|---|---|---|---|---|---|---|---|
| (1) | (2) | (3) | (4) | (5) | (6) | (7) | (8) | (9) |
| J021934.93−002432.3 Beam 17 ON | Mrk 592 | 02:19:41.2 | −00:15:20.66 | 1.6 | 9.2 | 0.026 | 290 | −10 |
| J043242.28−042257.0 Beam 13 ON | NGC 1607 | 04:32:03.1 | −04:27:35.6 | −9.8 | −4.6 | 0.016 | 217 | −39 |
| J043242.28−042257.0 Beam 07 ON | NGC 1611 | 04:33:05.9 | −04:17:50.7 | 5.9 | 5.1 | 0.014 | 136 | 175 |
| J043242.28−042257.0 Beam 12 OFF | | 04:31:33.2 | −04:32:54.6 | −17.3 | −10.0 | 0.013$^a$ | 328 | −262 |
| J043242.28−042257.0 Beam 18 ON | MCG-01-12-028 | 04:33:01.0 | −04:11:18.9 | 4.7 | 11.6 | 0.015 | 229 | 683 |
| J043242.28−042257.0 Beam 19 ON | NGC 1611 | 04:33:05.9 | −04:17:50.7 | 5.9 | 5.1 | 0.014 | 136 | 163 |
| J080028.55+152711.2 Beam 09 ON | AGC 188807 | 08:01:04.8 | +15:22:45.0 | 9.1 | −4.4 | 0.016 | 195 | 242 |
| J114040.73+594850.5 Beam 07 ON | NGC 3809 | 11:41:16.1 | +59:53:08.8 | 8.8 | 4.3 | 0.011 | 87 | 136 |
| J115704.07+221845.5 Beam 11 ON | | 11:57:04.1 | +22:08:49.1 | 0.0 | −9.9 | 0.023$^a$ | 275 | −87 |
| J115704.07+221845.5 Beam 16 ON | LEDA 5075770 | 11:56:32.8 | +22:27:41.5 | −7.8 | 8.9 | 0.022 | 306 | −343 |
| J121042.46+131848.4 Beam 12 ON | | 12:10:19.0 | +13:08:50.8 | −5.9 | −10.0 | 0.023$^a$ | 323 | 38 |
| J125815.27+272752.9 Beam 11 ON | TT 33 | 12:58:18.6 | +27:18:39.0 | 0.8 | −9.2 | 0.025 | 277 | −136 |
| J155206.94+410105.4 Beam 10 ON | SDSS J155221.79+405136.5 | 15:52:21.8 | +40:51:36.6 | 3.7 | −9.5 | 0.026 | 307 | −55 |
| J155206.94+410105.4 Beam 11 ON | SDSS J155221.79+405136.5 | 15:52:21.8 | +40:51:36.6 | 3.7 | −9.5 | 0.026 | 307 | 23 |
| J162758.43+390704.3 Beam 12 ON | SDSS J162728.89+385723.1 | 16:27:28.9 | +38:57:23.2 | −7.4 | −9.7 | 0.028 | 376 | −59 |
| J162812.10+404719.0 Beam 12 ON | | 16:27:42.0 | +40:37:21.4 | −7.5 | −10.0 | 0.030$^a$ | 413 | 326 |
| J162812.10+404719.0 Beam 17 ON | LEDA 2173685 | 16:28:17.6 | +40:58:01.2 | 1.4 | 10.7 | 0.029 | 376 | 77 |
| J162843.38+403219.0 Beam 13 ON | | 16:27:58.2 | +40:27:19.6 | −11.3 | −5.0 | 0.030$^a$ | 353 | 418 |

(1) Name of the central target and the number of FAST beam which the H I line detected from. The "ON/OFF" tag labels the ON/OFF-source integration when the signal was detected. See Section 3.1 for more details about the observing mode, (2) name of the identifier from NED. Sources with no identifier are left empty, (3) & (4) central position of the beam, (5) & (6) spatial offset to the central target in RA and DEC, (7) redshift of the identifier. For sources failed to match an identifier, we present the redshift converted from the heliocentric velocities, (8) projected distance to the central target. For sources with no identifiers confirmed, the distance is measured between the beam position and the central target, (9) radio heliocentric velocity offset to the central dwarf galaxy. When the targeted dwarf galaxy is not detected, the velocity offset is calculated between the radio velocity of the source and optical velocity of the central target.
$^a$Redshift converted from the heliocentric velocities.
$^b$GSWLC stellar mass.

dation No. 2021M703236. This work is supported by the International Partnership Program of Chinese Academy of Sciences, Program No. 114A11KYSB20210010, and National Key R&D Program of China No.2023YFE0110500. NKY acknowledges support by the projects funded by China Postdoctoral Science Foundation No. 2022M723175 and GZB20230766. This work made use of the data from FAST (Five-hundred-meter Aperture Spherical radio Telescope). FAST is a Chinese national mega-science facility, operated by National Astronomical Observatories, Chinese Academy of Sciences. This research made use of the NASA/IPAC Extragalactic Database [4], which is funded by the National Aeronautics and Space Administration and operated by the California Institute of Technology. This research made use of Astropy, a community-developed core Python package for Astronomy (Astropy Collaboration et al. 2013, 2018). YC thanks the Center for Astronomical Mega-Science, Chinese Academy of Sciences, for the FAST distinguished young researcher fellowship (19-FAST-02) and the support from the National Natural Science Foundation of China (NSFC) under grant No. 12050410259 and the Ministry of Science and Technology (MOST) of China grant no. QNJ2021061003L.

## DATA AVAILABILITY

The data underlying this article are available in the article.

[4] http://ned.ipac.caltech.edu

## APPENDIX A: FAST SPECTRA

In this Appendix, we present the measured H I line spectra of the targets (Figure A1) and their 19-beam H I spectra overlaid on SDSS DR16 optical images (Figure A2). Please see Section 4.1 and Section 5.2 for more detailed descriptions and discussions.

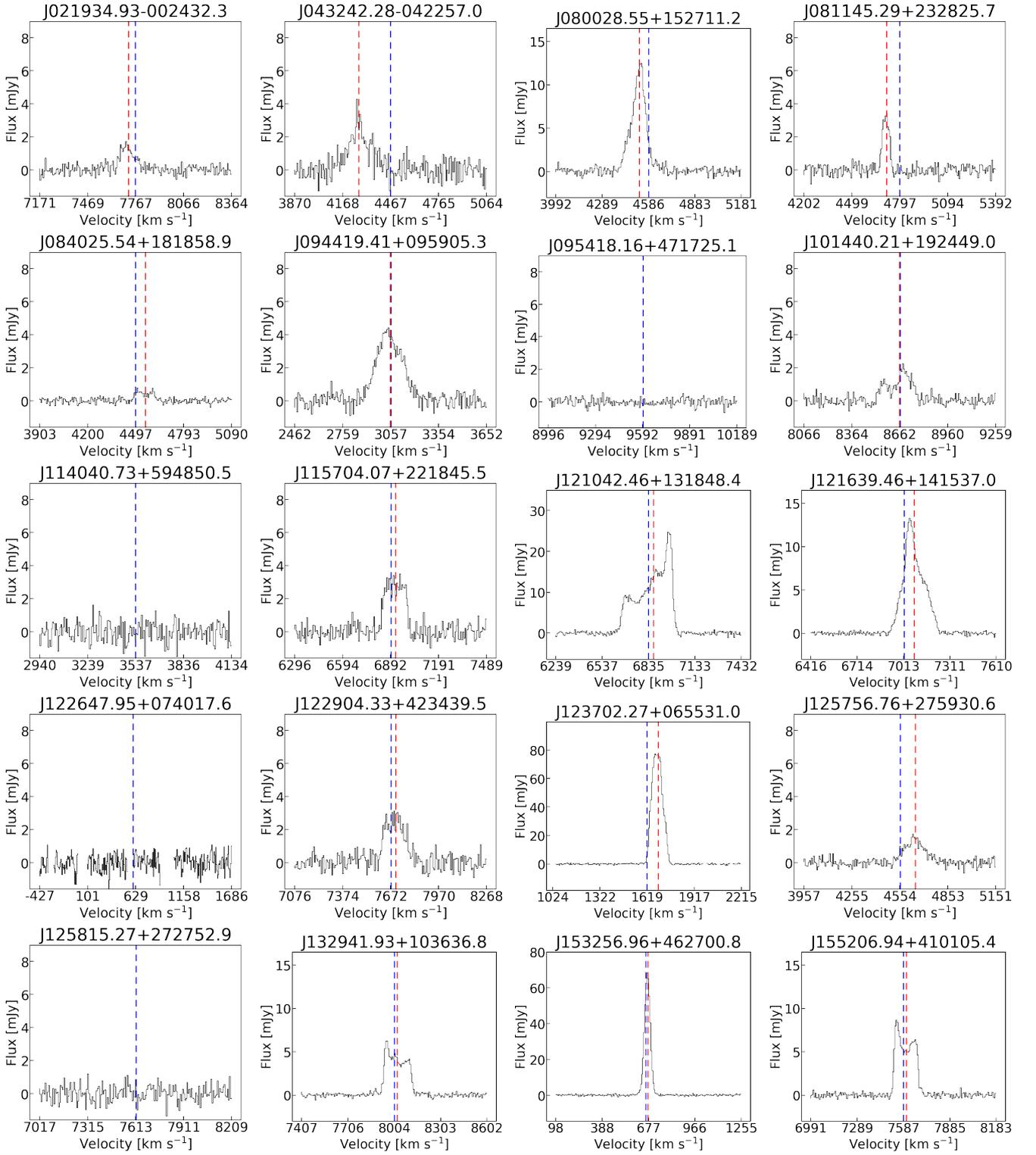

**Figure A1.** H I spectrum of the targets. The spectrum are stacked over the central beam (Beam 01) and the outer beam (Beam 08 or 14). The blue lines label the optical velocity calculated from redshift. For the detected sources, the red vertical lines denote the central heliocentric velocity of the H I signal.

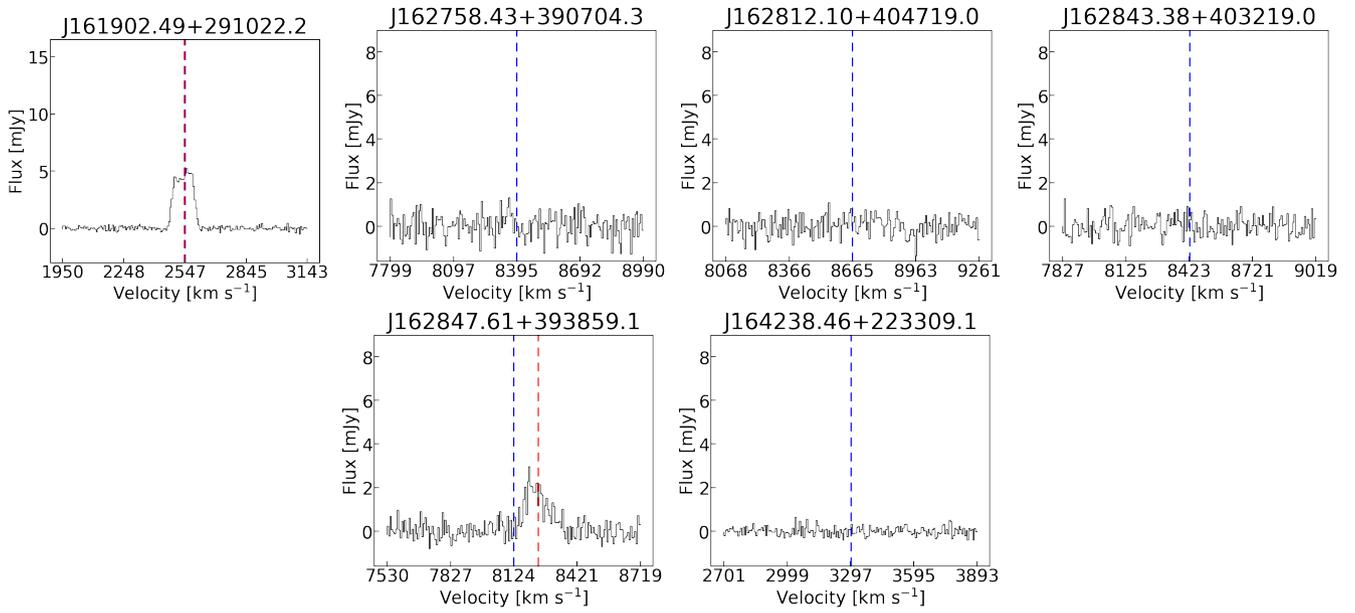

**Figure A1.** Continued.

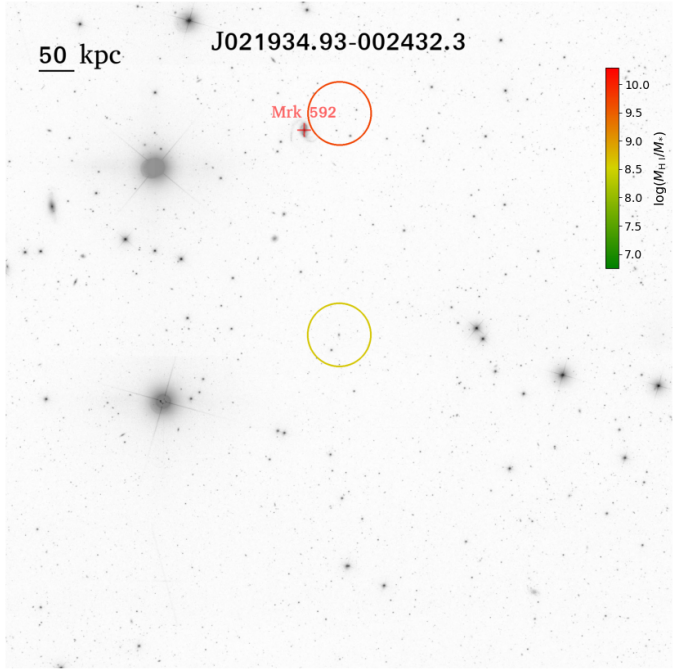
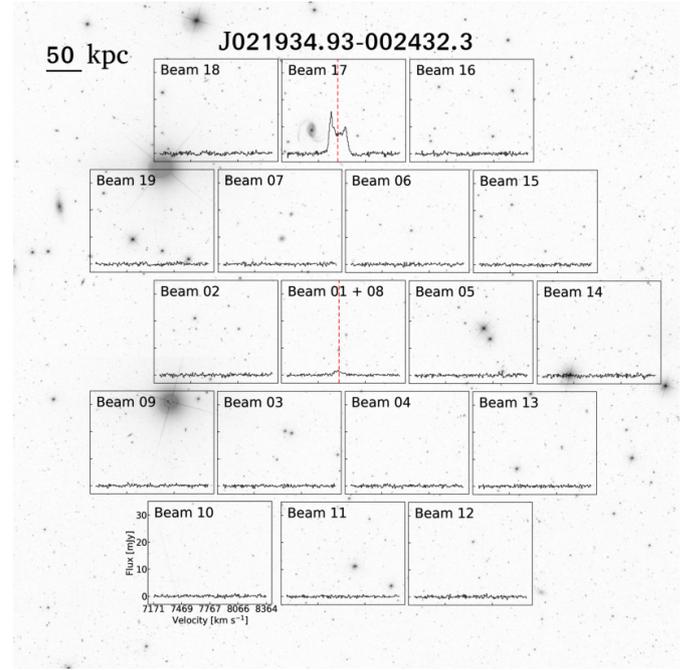
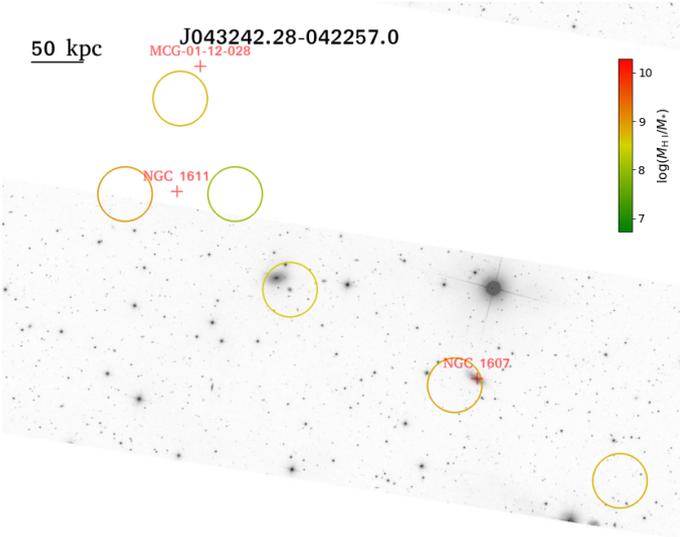
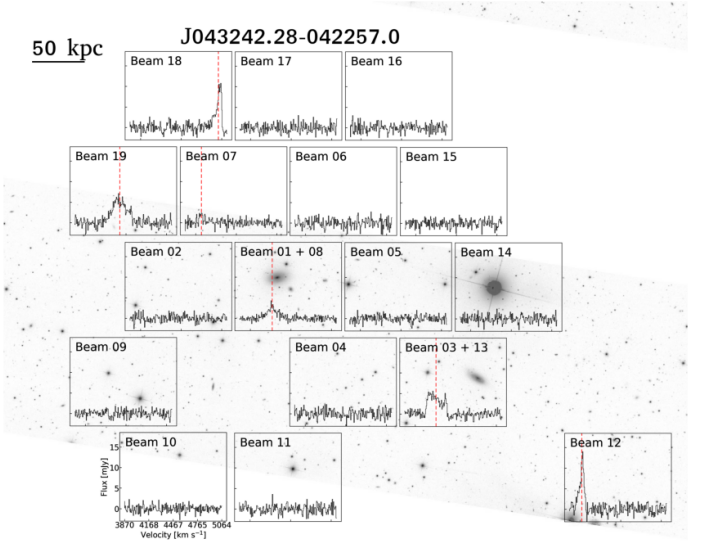

**Figure A2.** H I spectra observed with FAST 19-beam, overlaid on SDSS DR 16 optical images. During the observations, ON-source observations were done with the central beam of the FAST 19 beam pointing at the targeted dwarf galaxies. OFF-source observations were made by shifting the outer most beam (beam 08 or 14) to the targeted object. As a result, for each source, 38 (19 × 2) datasets will be recorded. There are 9 positions have two redundant observations from different beams during ON and OFF-source observations, while 29 positions are only observed by single beam. When an H I detection is observed by two different beams, we plot the spectrum at the location where the H I signal is emitted, and omit the plot of the beam that is used as background reference. The circles on the left panel locate the FAST beam positions of detected signals, they are colored by the detected H I mass. Known identifiers to the H I sources surrounding the central dwarf galaxies are labeled by the red crosses. The 19-beam H I spectra are displayed on the right panels. Duplicated spectra are stacked and plotted on top of the on-source beam position. The red dashed lines illustrate the central velocities of the detected H I emission lines. The black bar on the upper left corner illustrate the projected scale of 10 or 50 kpc in the sky.

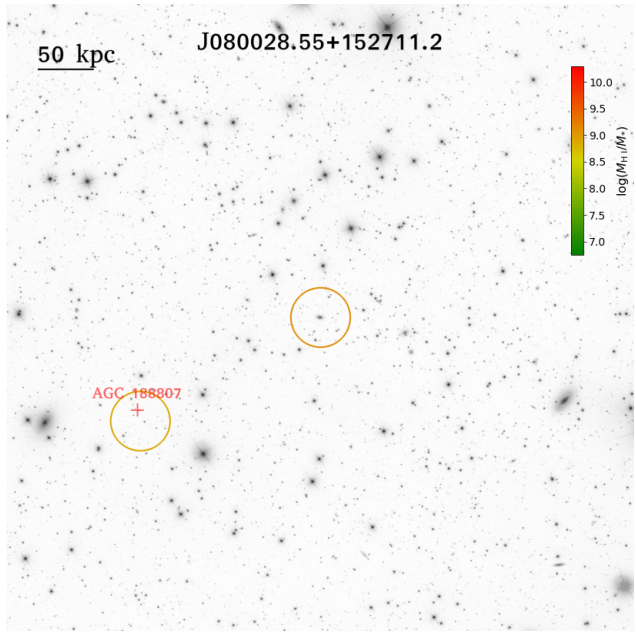
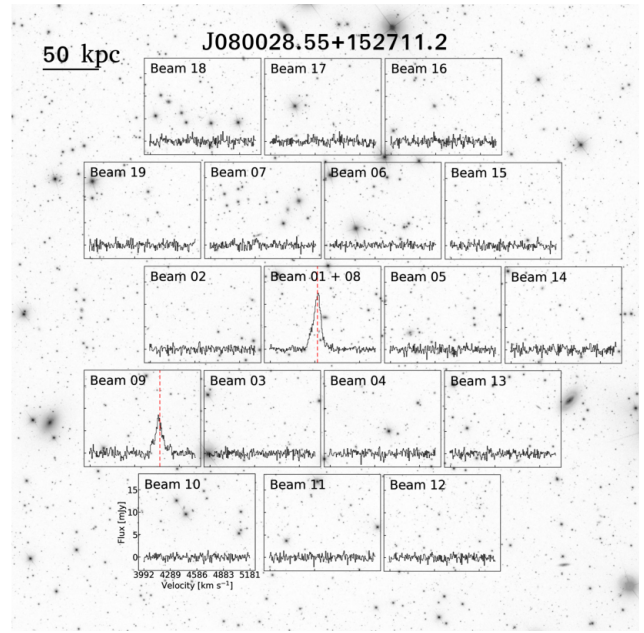
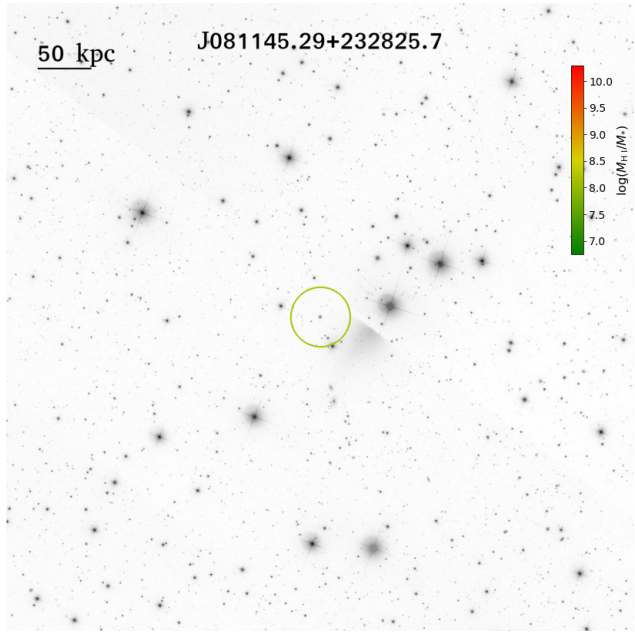
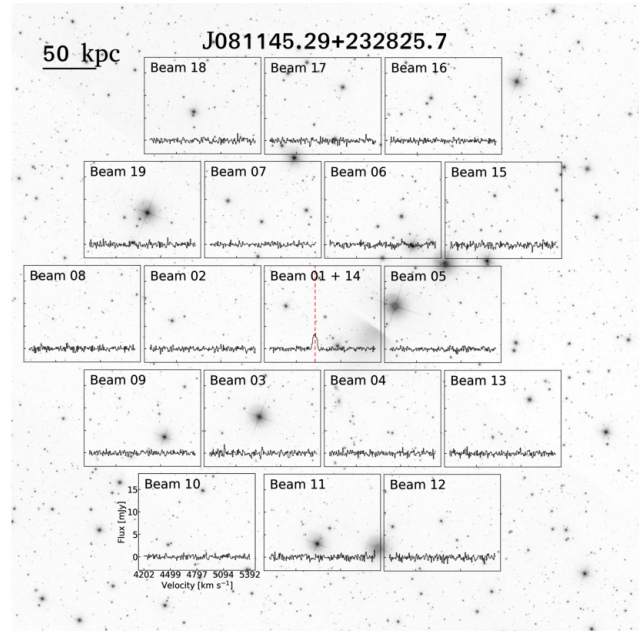
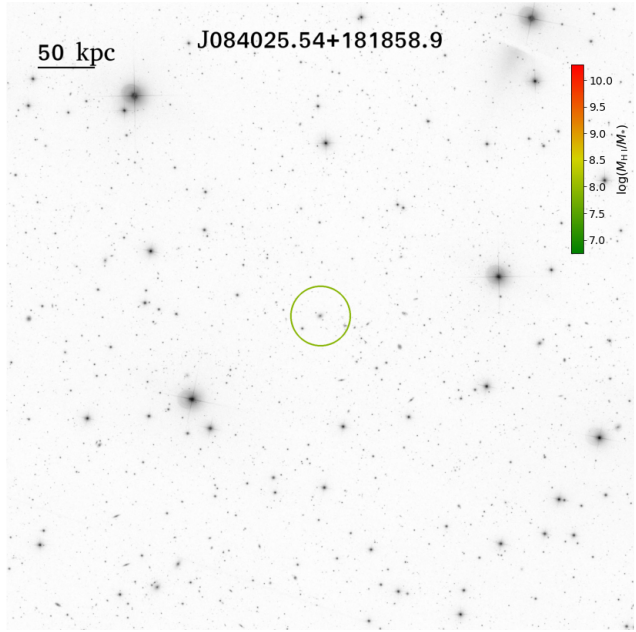
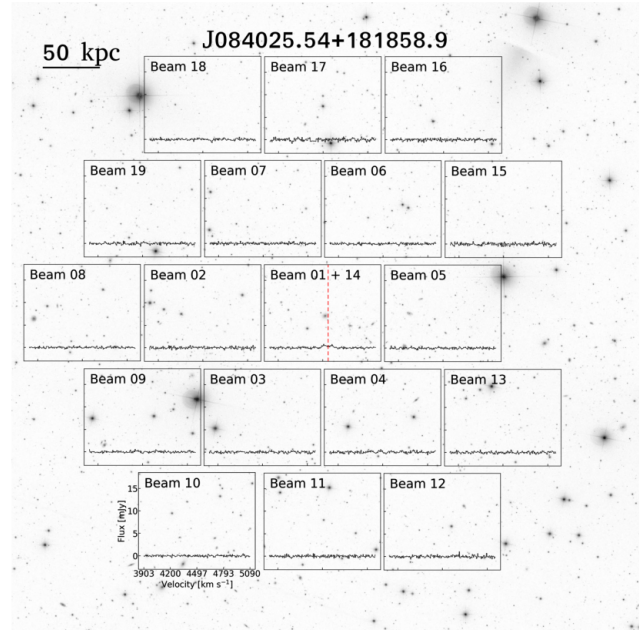

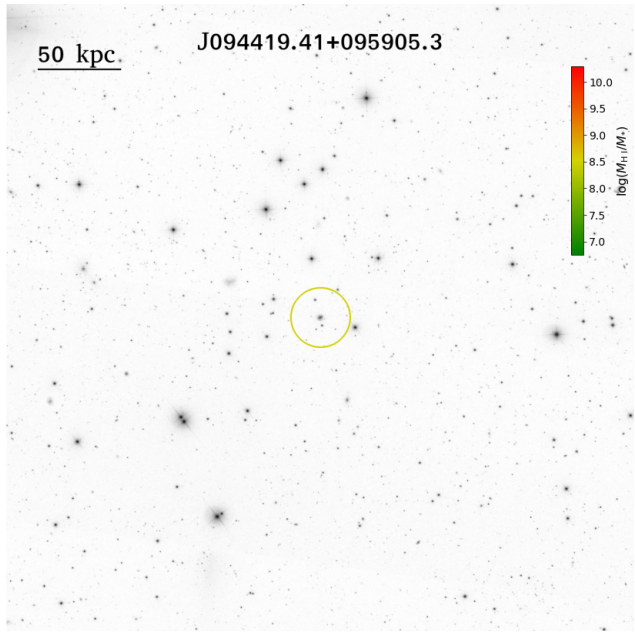
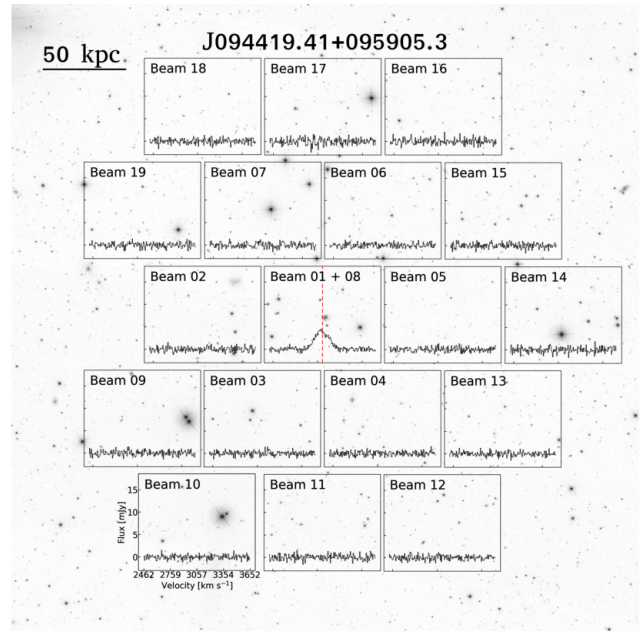
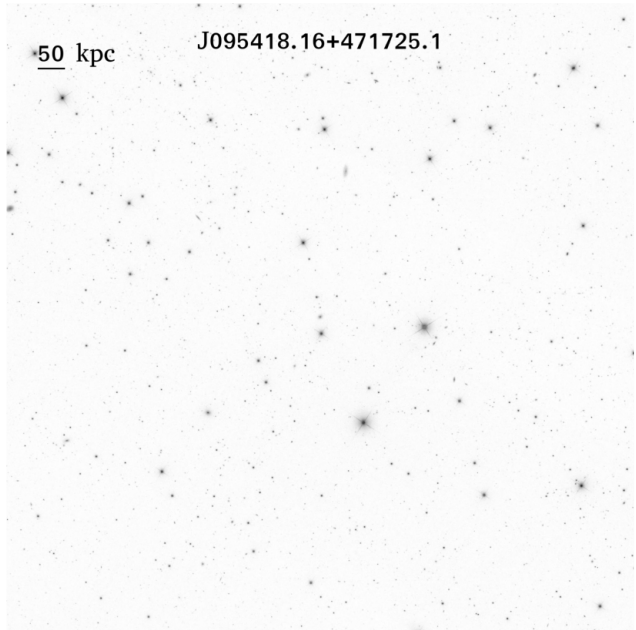
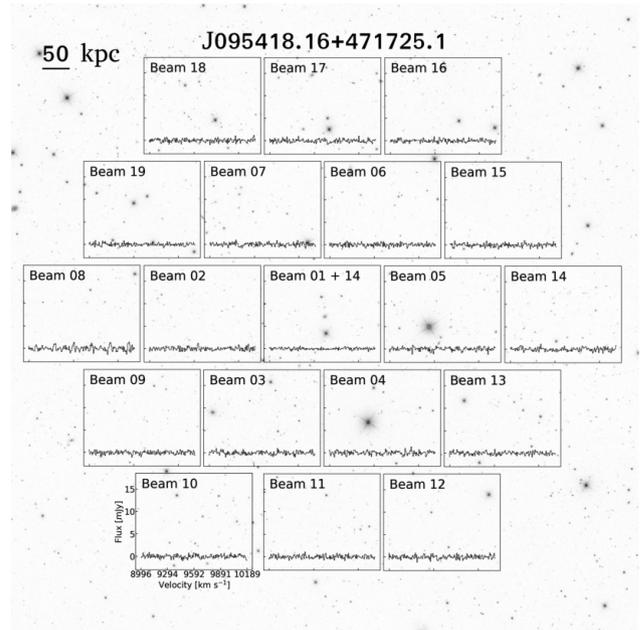
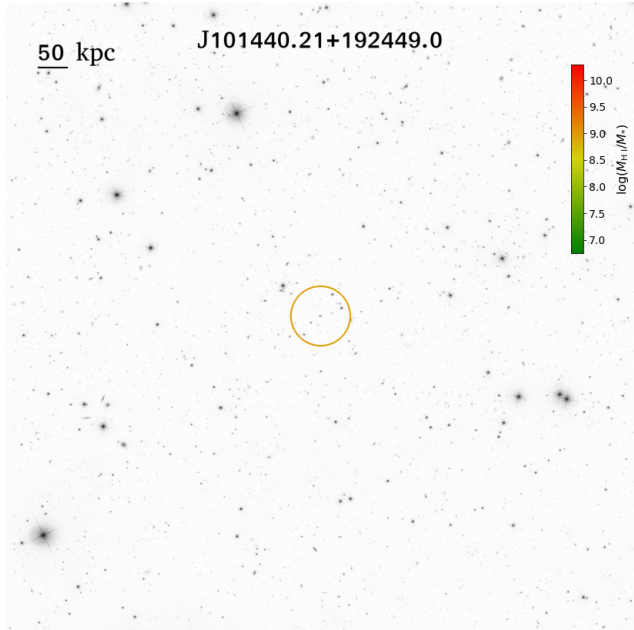
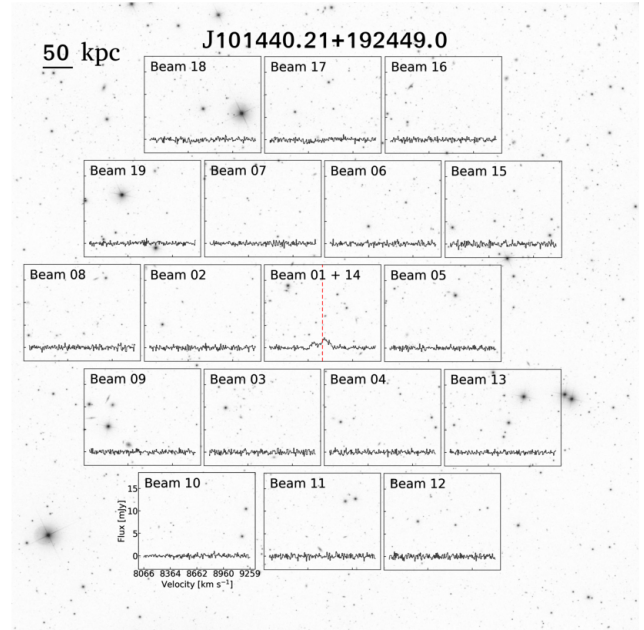

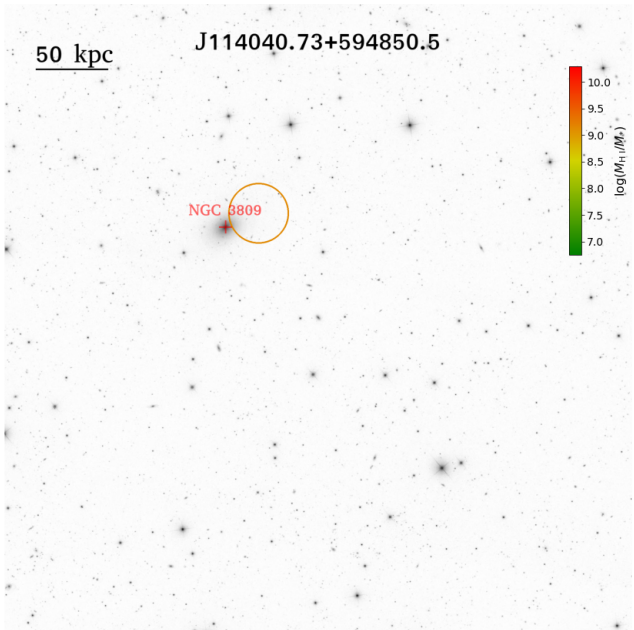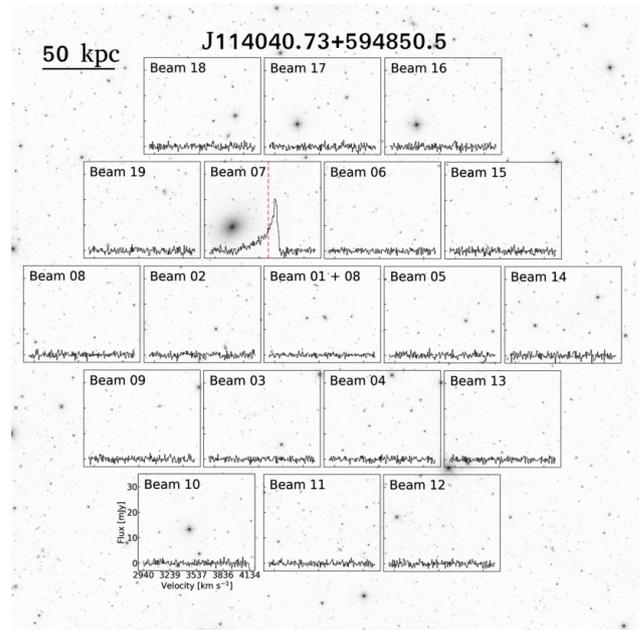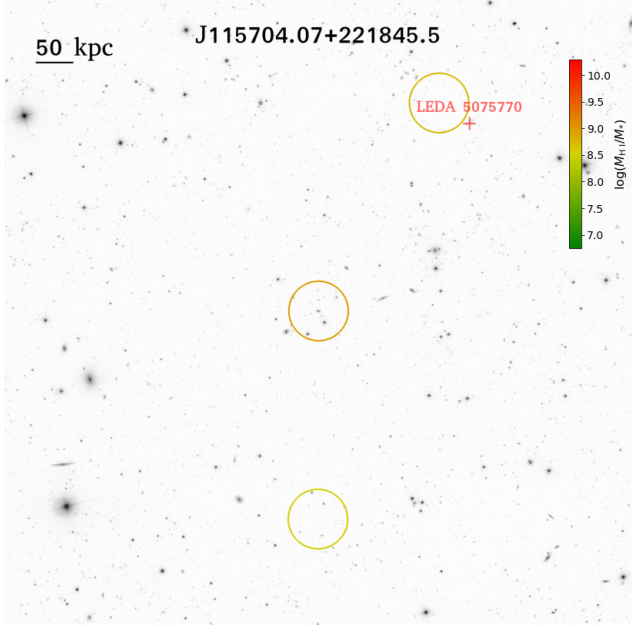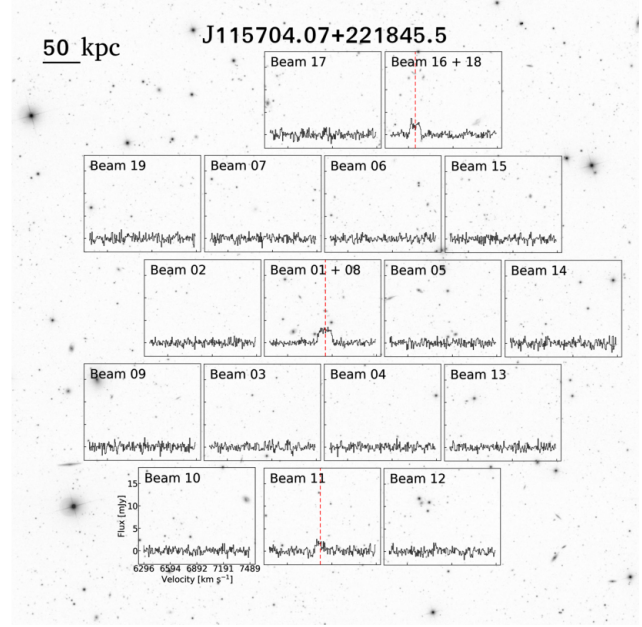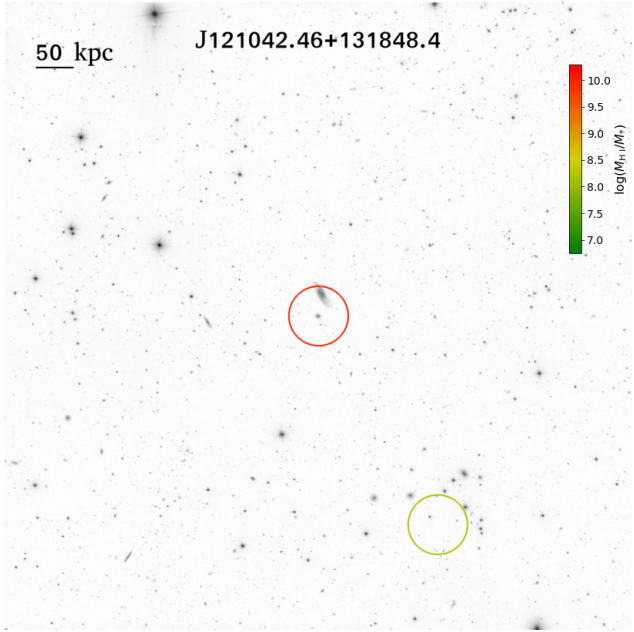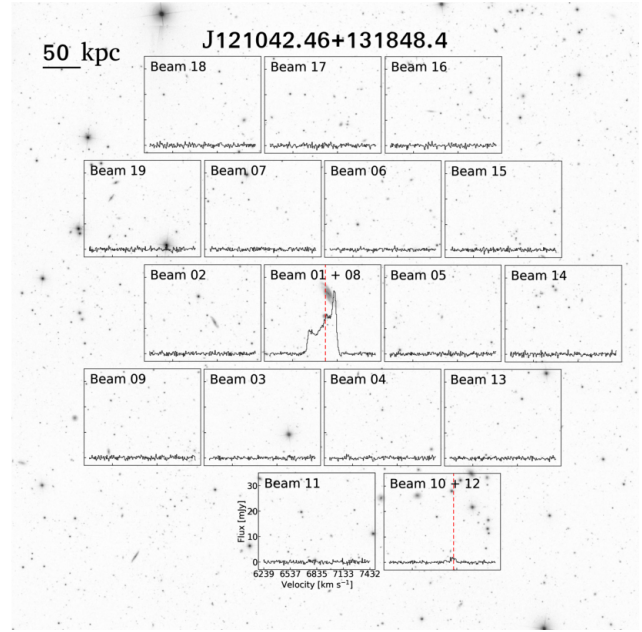

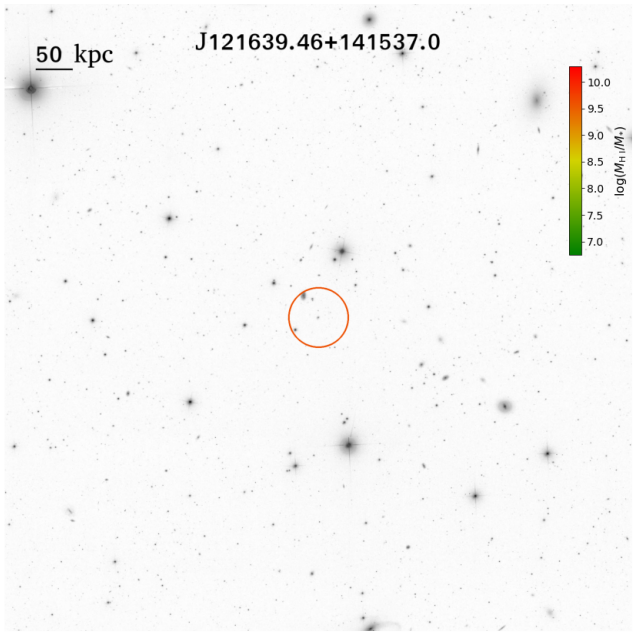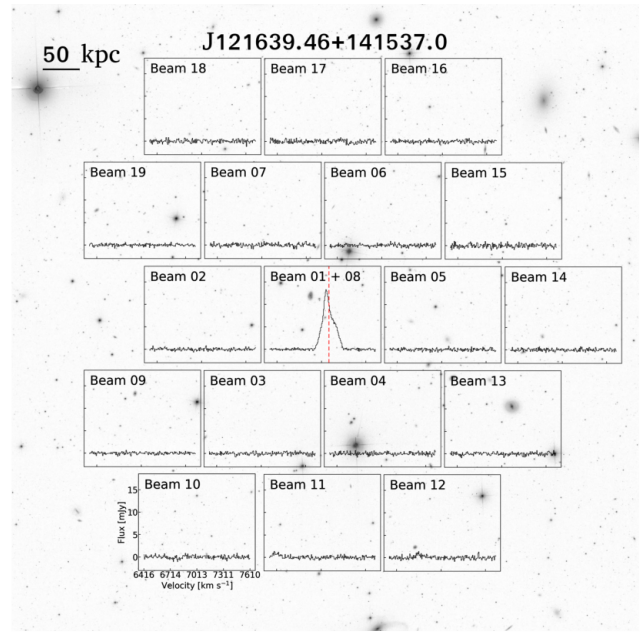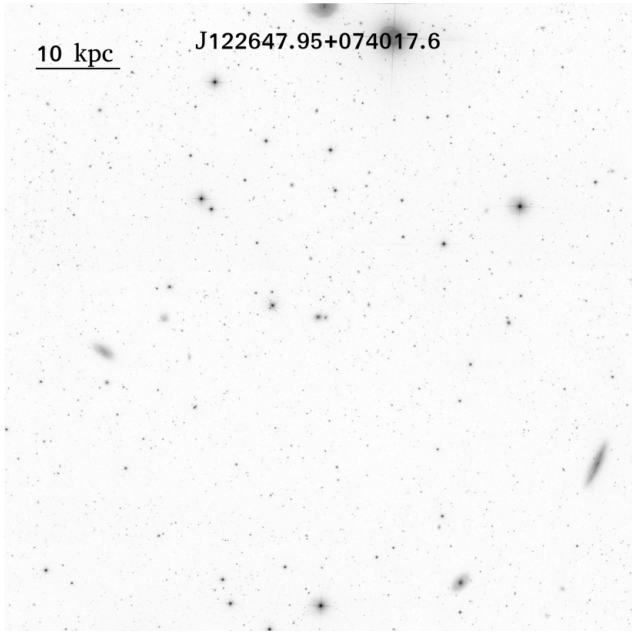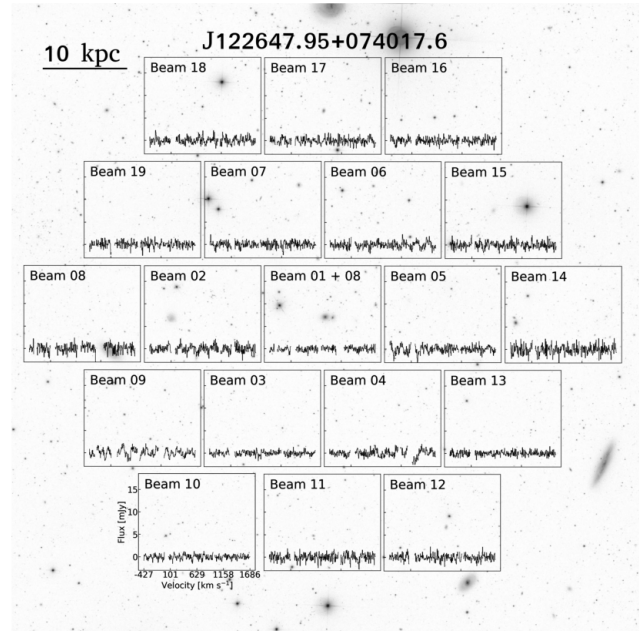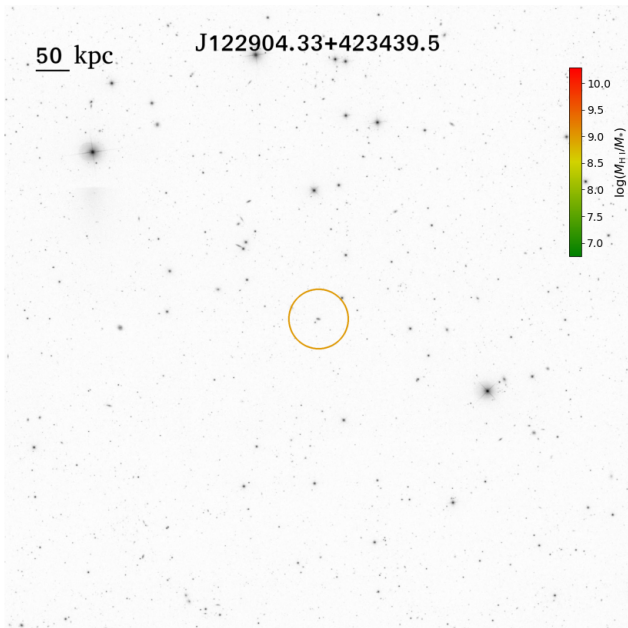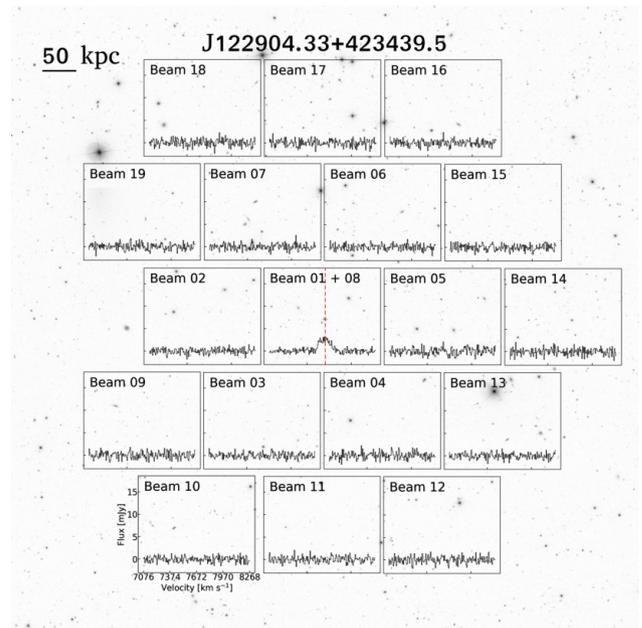

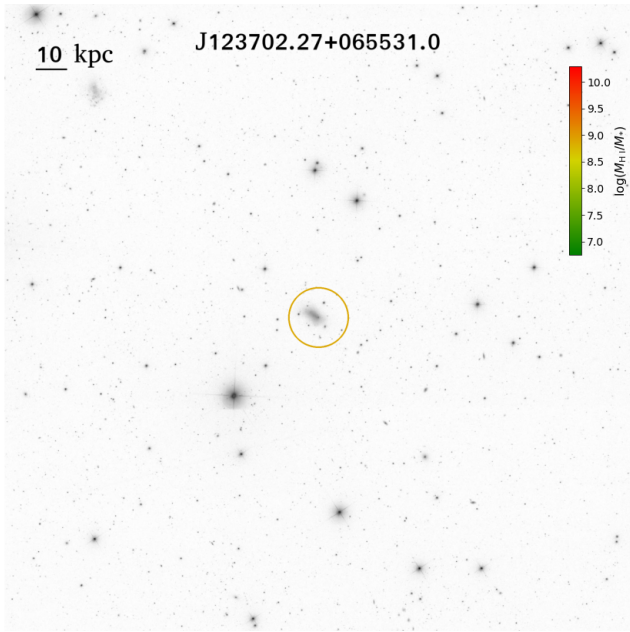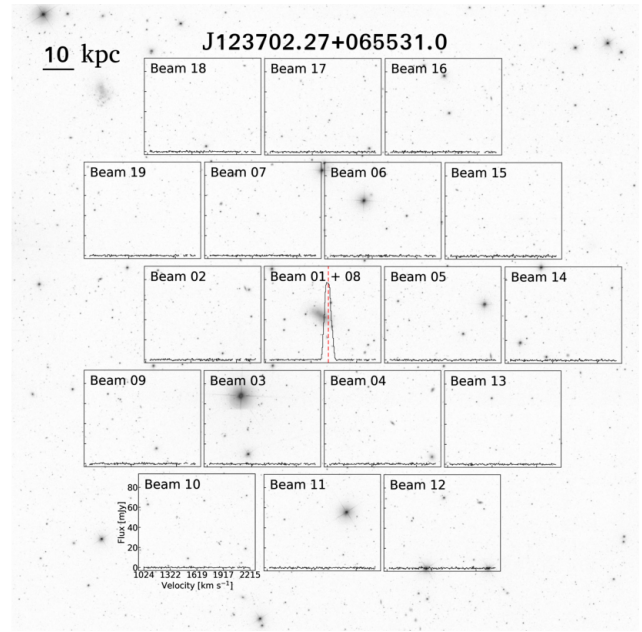
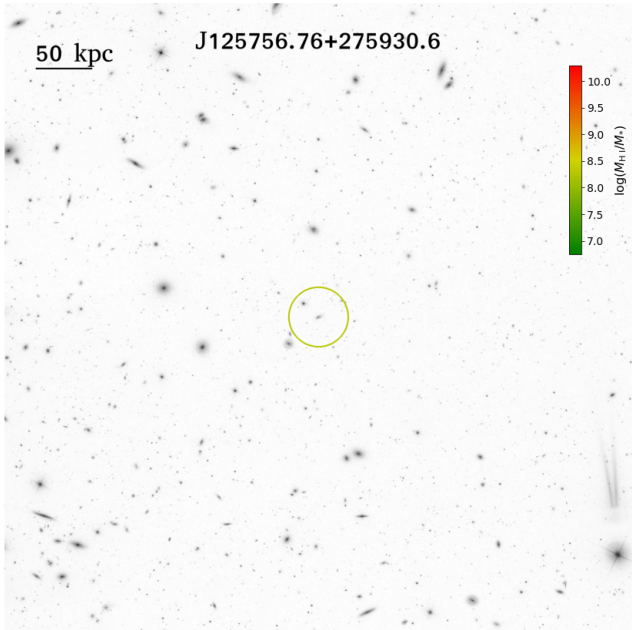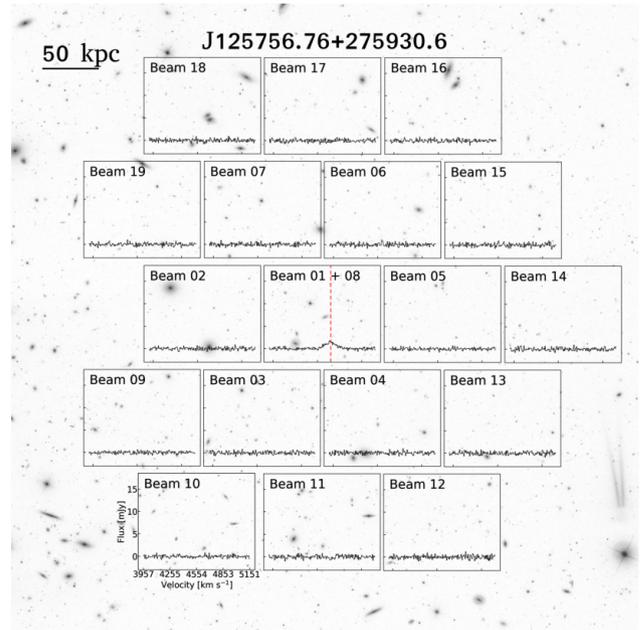
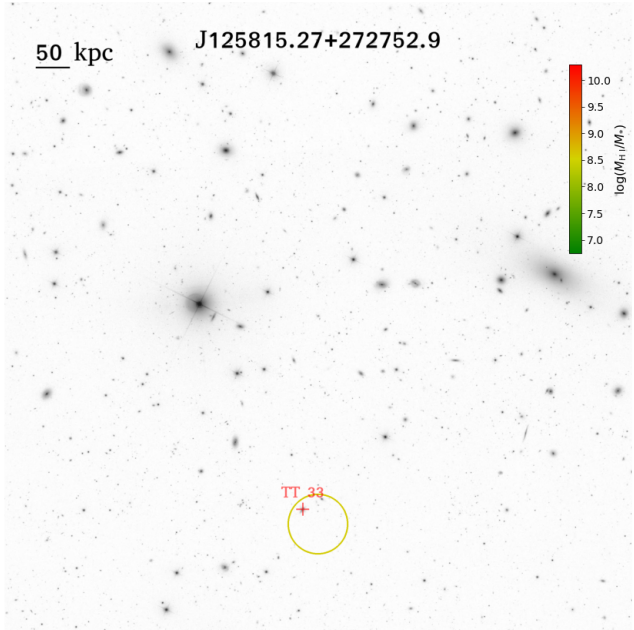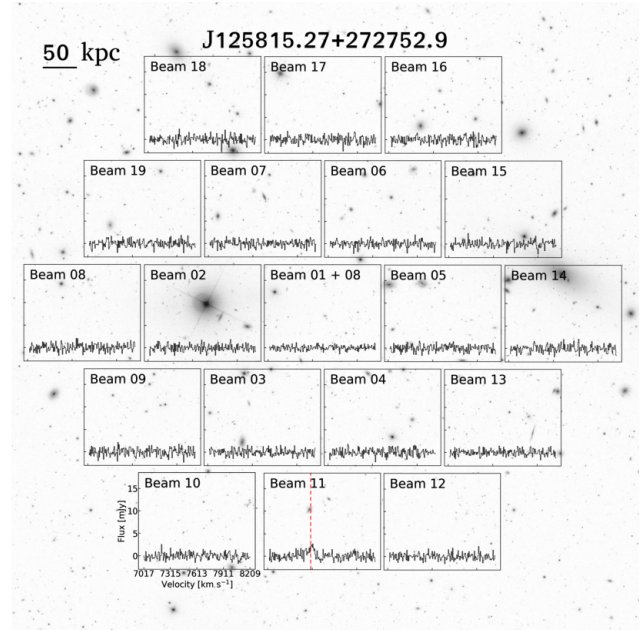

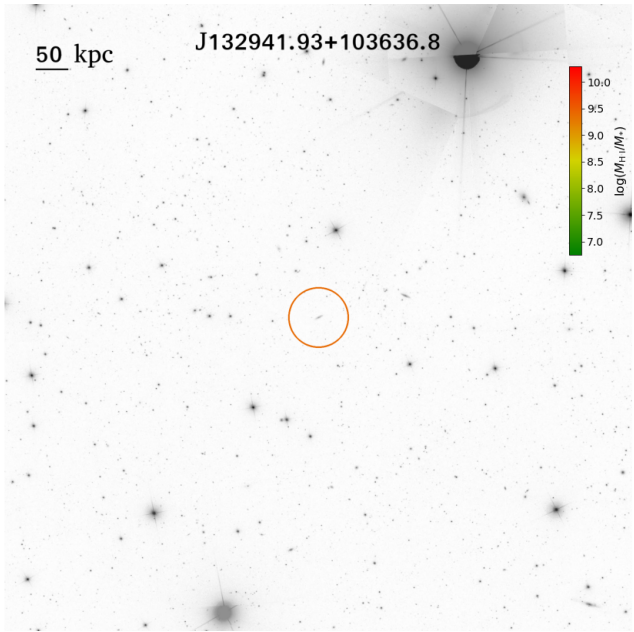
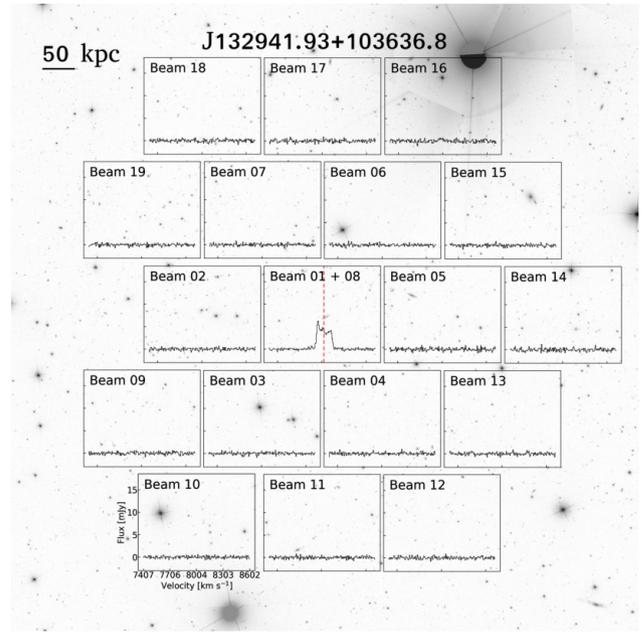
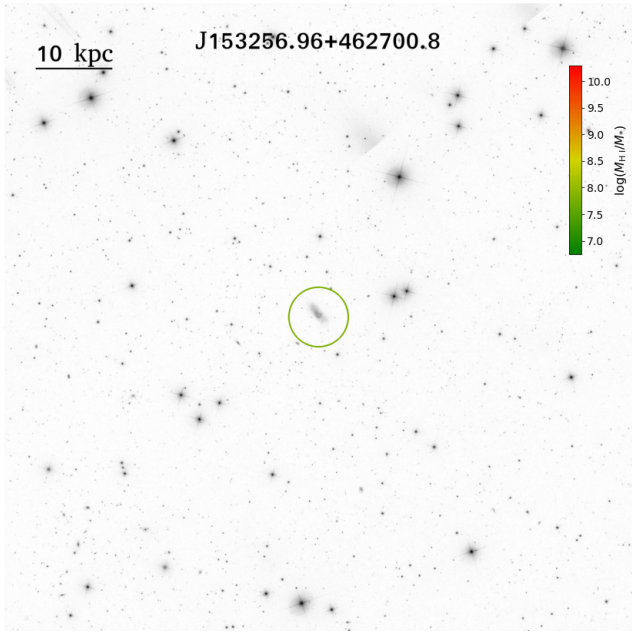
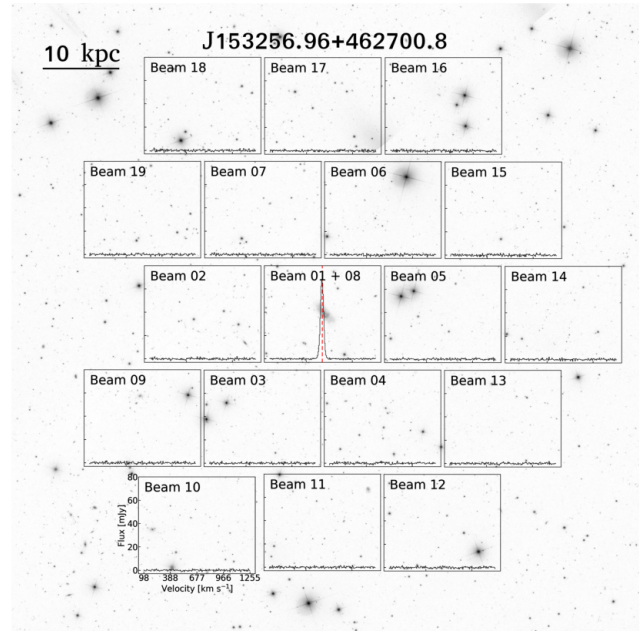
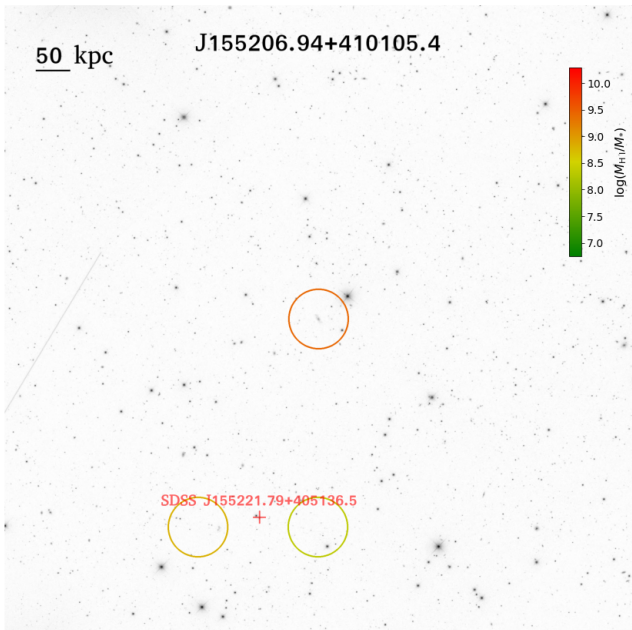
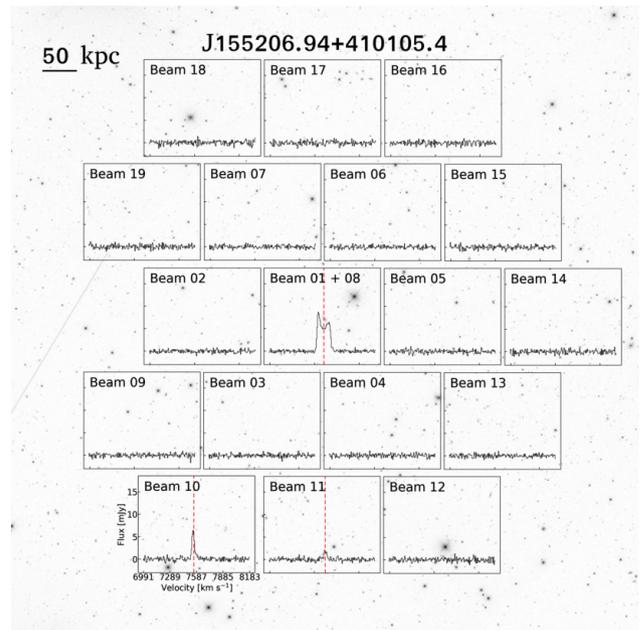

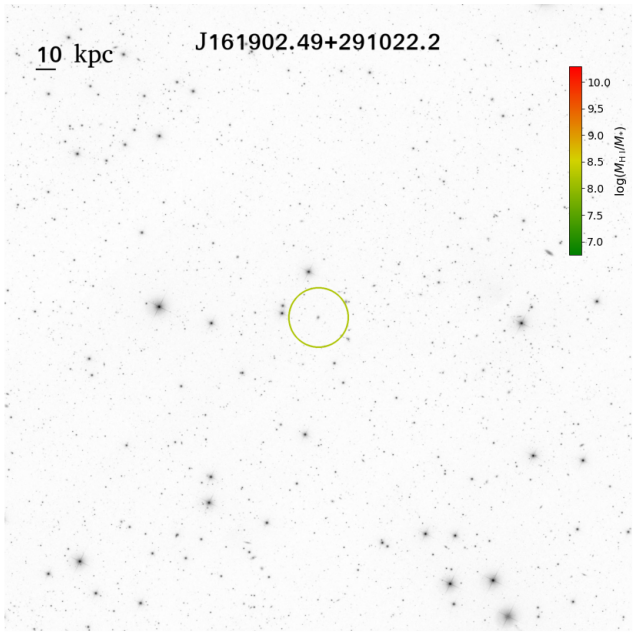
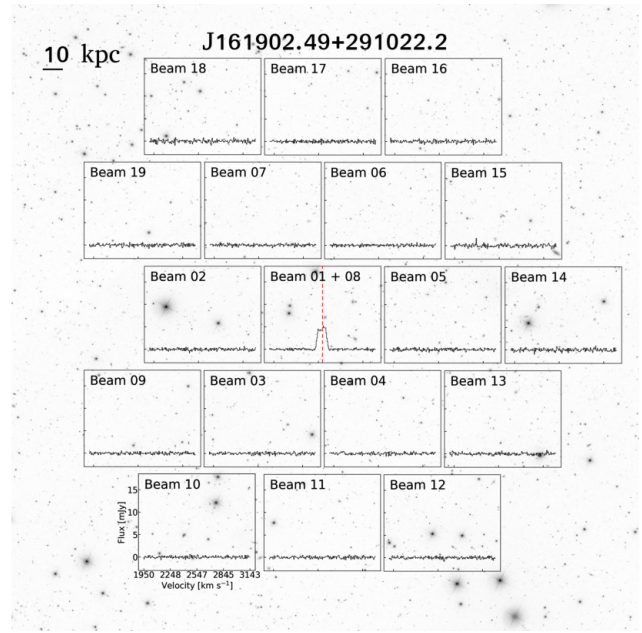
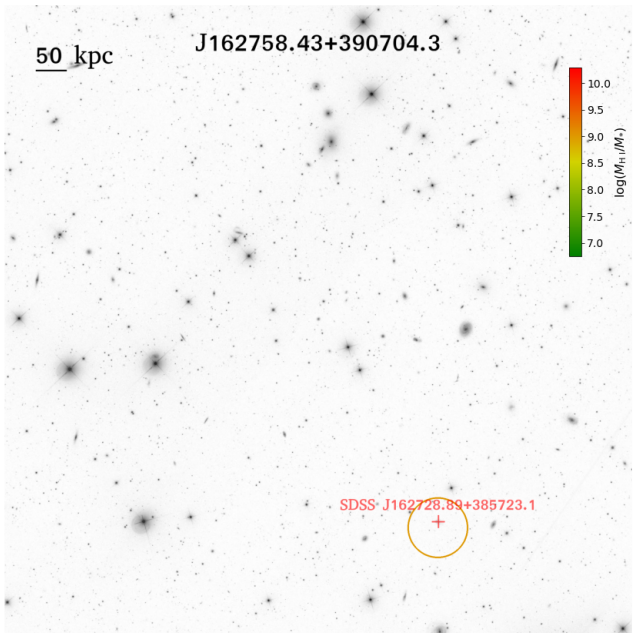
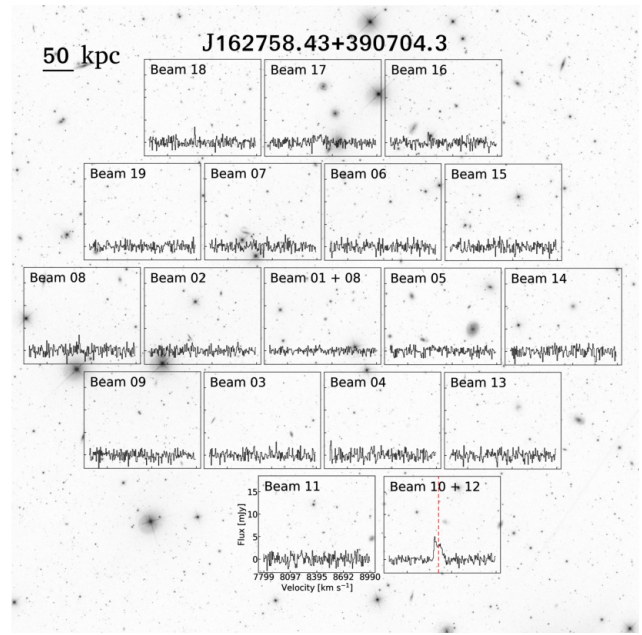
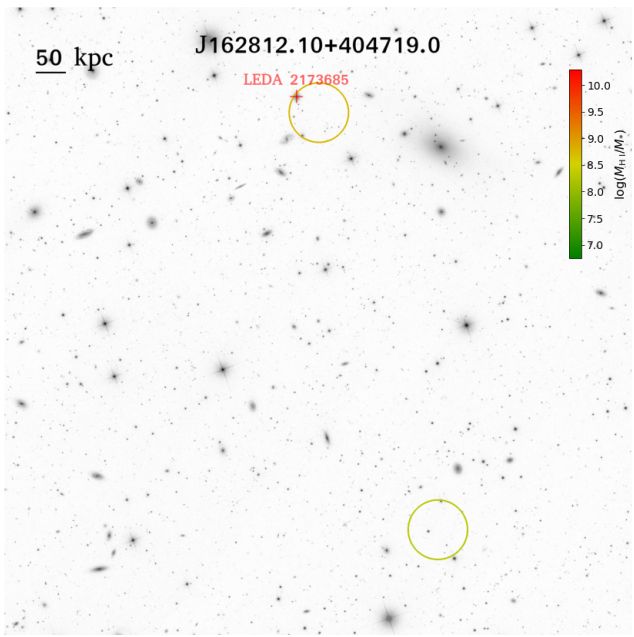
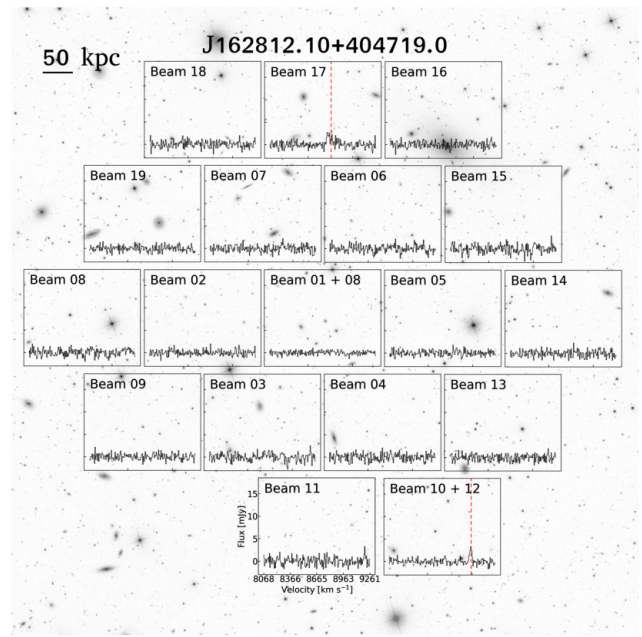

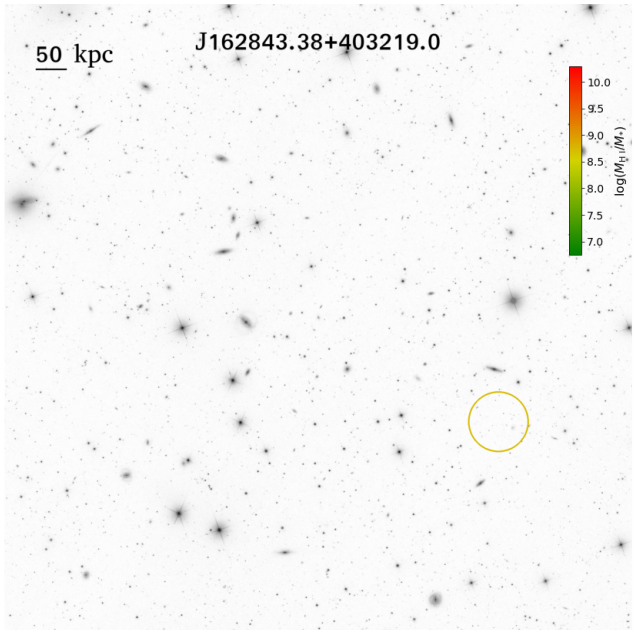
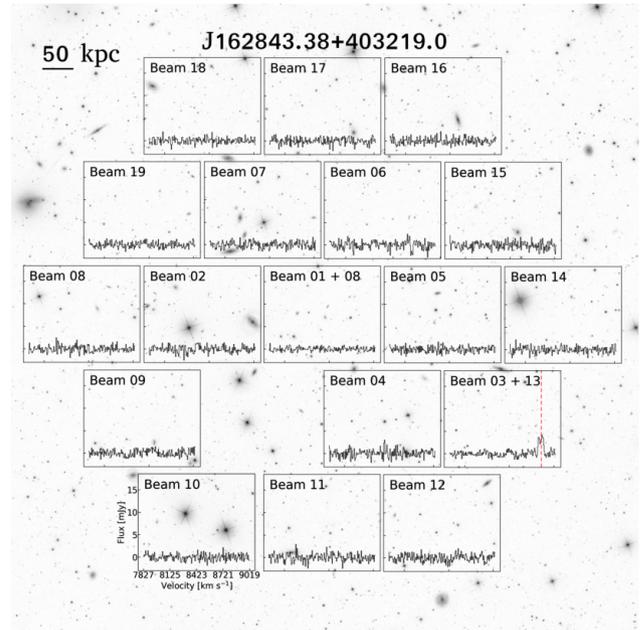
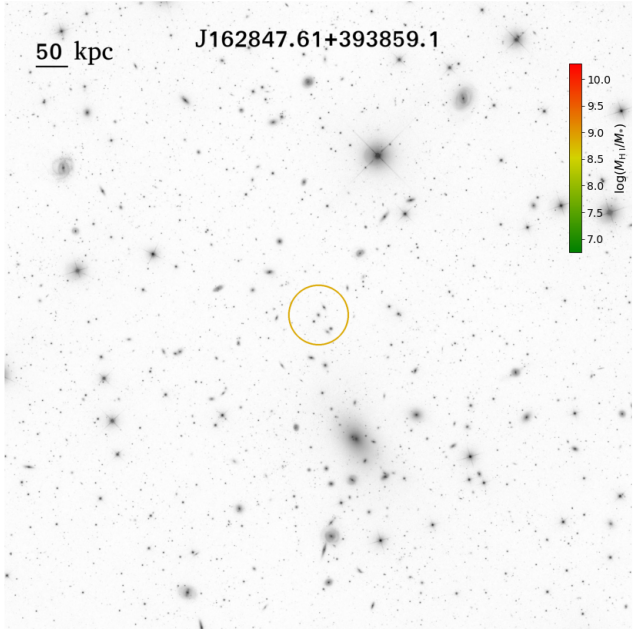
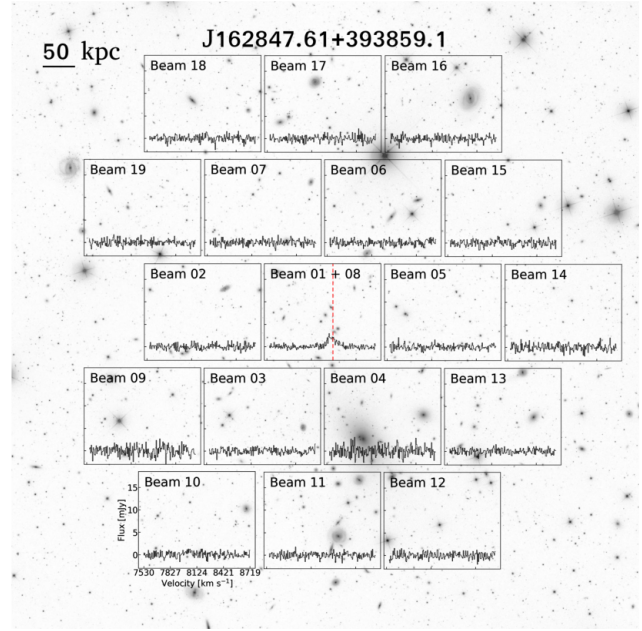
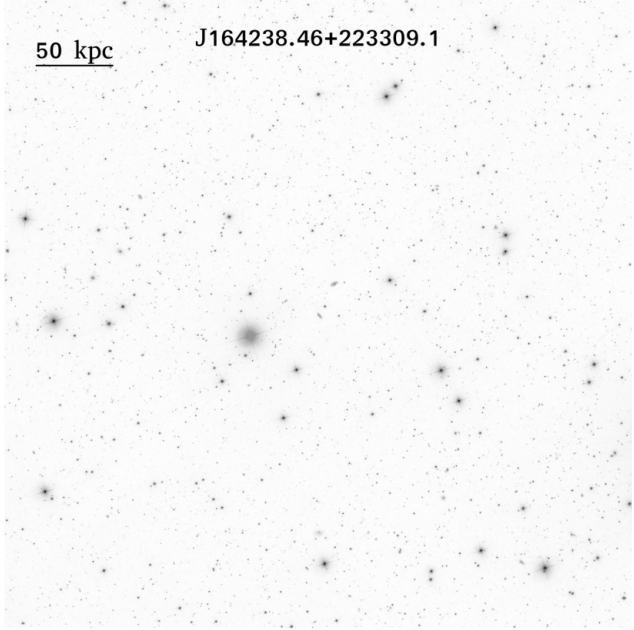
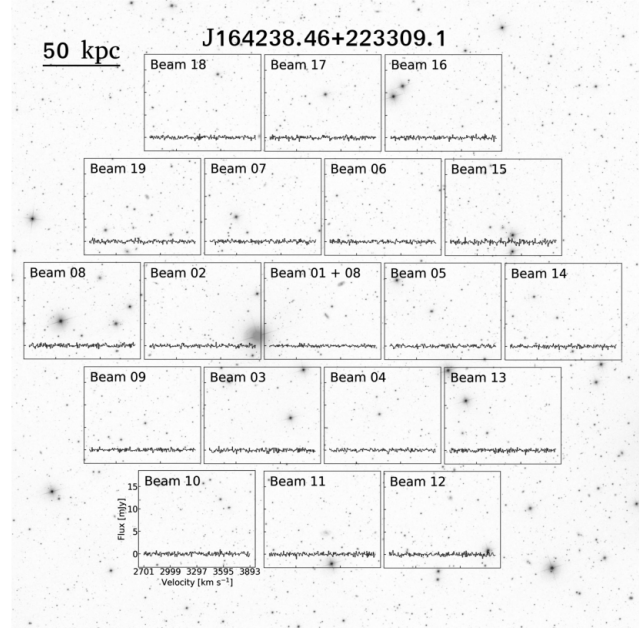